\documentclass[11pt,a4paper]{article}
\usepackage{amsmath,amssymb}
\usepackage{yfonts}
\usepackage{bbm}
\usepackage{cancel}
\usepackage{rotating}
\usepackage{cite}

\def\hybrid{\topmargin -20pt    \oddsidemargin 0pt
        \headheight 0pt \headsep 0pt
        \textwidth 6.25in       
        \textheight 9 in       
        \marginparwidth .875in
        \parskip 5pt plus 1pt
          \jot = 1.5ex
   }
\hybrid
\numberwithin{equation}{section}
\numberwithin{table}{section}

\newcommand{\be}{\begin{equation}}
\newcommand{\ee}{\end{equation}}
\newcommand{\bea}{\begin{eqnarray}}
\newcommand{\eea}{\end{eqnarray}}

\newcommand{\un}{{\bf 1}}

\newcommand{\f}{{\bf 5}}
\newcommand{\fb}{{\bf \bar{5}}}
\newcommand{\te}{{\bf 10}}
\newcommand{\teb}{{\bf \bar{10}}}
\newcommand{\op}{\oplus}

\newcommand{\tw}{\text{w}}
\newcommand{\tu}{\text{u}}
\newcommand{\tv}{\text{v}}

\newcommand{\TM}{\mathbf{10}_\mathrm{M}}
\newcommand{\FU}{\mathbf{5}_{\mathrm{H}_\mathrm{u}}}
\newcommand{\FD}{\bar{\mathbf{5}}_{\mathrm{H}_\mathrm{d}}}
\newcommand{\FM}{\bar{\mathbf{5}}_\mathrm{M}}
\newcommand{\OO}{\mathbf{1}}
\newcommand{\ON}{\mathbf{1}_{\nu_\mathrm{R}}}

\newcommand{\mc}{\mathcal}

\begin{document}
\thispagestyle{empty}
\baselineskip=14pt
\parskip 5pt plus 1pt

\vspace*{-1.5cm}
\begin{flushright}    
  {\small
  
  }
\end{flushright}

\vspace{2cm}
\begin{center}        
  {\LARGE On $E_8$ and F-Theory GUTs}
\end{center}

\vspace{0.75cm}
\begin{center}        
  Florent Baume$^1$, Eran Palti$^1$, Sebastian Schwieger$^{1,2}$
\end{center}

\vspace{0.15cm}
\begin{center}        
 ${^1}$ Institut f\"ur Theoretische Physik, Ruprecht-Karls-Universit\"at, \\
             Philosophenweg 19, 69120, Heidelberg, Germany
             \\[0.15cm]
 ${^2}$ Instituto de F\'{i}sica Te\'{o}rica UAM/CSIC, Universidad Aut\'{o}noma de Madrid, \\
				Cantoblanco, 28049 Madrid, Spain
        \\[0.15cm]
\end{center}

\vspace{2cm}


\begin{abstract}
  \vspace{0.5cm} 
	We study correlations between the massless field spectra of F-theory fibrations supporting an $SU(5)$ gauge symmetry extended by Abelian symmetries and the spectra that arise from the group $E_8$. The adjoint representation of $E_8$ leads to six different classes of matter spectra upon Higgsing $E_8$ to $SU(5)\times U(1)^n$. Of 27 different smooth F-theory elliptic fibrations constructed in the literature, satisfying certain genericness and flatness criteria, the matter spectrum of only one can be embedded in these six theories, thereby apparently ruling out any connection. We define an extension of the spectrum arising from the adjoint of $E_8$ by introducing new $SU(5)$-singlet fields with Abelian charges such that there exists a cubic gauge invariant coupling between any three representations. Higgsing by these new singlets leads to a further 20 classes of spectra, and we find that all the F-theory fibrations can then be embedded in this extended set. These results show that $E_8$, when extended in this specific way, may still have a role to play in controlling the possible matter spectra in F-theory. We give an explicit geometric example of the presence of the extending singlets and their Higgsing. We discuss some phenomenological applications of the new set of theories, in particular due to the existence of a ${\mathbb Z}_2$ matter parity. Finally we make some comments on the Heterotic duals of the F-theory fibrations which extend $E_8$ in this way.

\end{abstract}
\vspace{1.7cm}

\clearpage


\newpage

\setcounter{page}{1}

\section{Introduction}

F-theory provides a geometric formulation of various aspects of non-perturbative type IIB string theory \cite{Vafa:1996xn,Morrison:1996na,Morrison:1996pp}. It is a framework which is particularly appropriate for studying the realisation of Grand Unified Theories (GUTs) in String Theory, see \cite{Weigand:2010wm,Heckman:2010bq,Maharana:2012tu} for reviews. From a IIB perspective these arise on the intersections of 7-branes, while in the geometric F-theory formulation they correspond to an intricate singularity structure on Calabi-Yau (CY) manifolds. The minimal GUT group is $SU(5)$ and there has been much work in recent years on constructing F-theory geometries which exhibit an $SU(5)$ gauge group that is extended by some further Abelian symmetries \cite{Grimm:2010ez,Krause:2011xj,Krause:2012yh,Grimm:2011fx,Mayrhofer:2012zy,Braun:2013yti,Borchmann:2013jwa,Cvetic:2013nia,Grimm:2013oga,Braun:2013nqa,Cvetic:2013uta,Cvetic:2013jta,Borchmann:2013hta,Bies:2014sra,Braun:2014pva,Kuntzler:2014ila,Klevers:2014bqa,Garcia-Etxebarria:2014qua,Mayrhofer:2014haa,Braun:2014qka,Lawrie:2014uya}. Although there are by now quite a few examples of such constructions we are still missing a systematic understanding of what are the possible symmetries and matter spectra that can be realised in such models. This can be contrasted with early F-theory model building where a local approach was used to build geometries based on the spectral-cover construction \cite{Donagi:2008ca,Beasley:2008dc,Donagi:2009ra,Marsano:2009gv,Marsano:2009wr,Dudas:2010zb,Dolan:2011iu}. There the geometries could be described as Higgs bundles over the divisor supporting the GUT group where the Higgs took values inside the commutant of the $SU(5)$ GUT group in $E_8$, which is again $SU(5)$ due to the decomposition $E_8 \supset SU(5)_{GUT} \times SU(5)_{\perp}$. The possible symmetries and matter charges that can arise from such spectral cover constructions can be easily classified as they arise from Higgsing $E_8$ to $SU(5)$ using its adjoint representation. Of course there remained much data of the theory, such as the massless spectrum and the values of operators, which depends on the precise details of the background geometry and fluxes, but the possible Abelian symmetries for any such model were embeddable inside the group $E_8$ and the possible matter charges under the symmetries were all realised in the decomposition of a single ${\bf 248}$ adjoint representation of $E_8$.

In this paper, with the aim of moving towards a more systematic understanding of the set of possibilities for F-theory GUT constructions, we study if there is a similar role that the group $E_8$ can play in constraining the possible symmetries and matter charges in full global F-theory models. In considering such a possibility there are immediate restrictions on the role that $E_8$ can play. First we know that certainly the total gauge group of F-theory models can be much larger than $E_8$ and in fact can contain thousands of $E_8$ factors \cite{Candelas:1997eh}. However each non-Abelian gauge group will be localised on a separate divisor in the geometry and therefore in considering the matter spectrum on the specific $SU(5)_{GUT}$ divisor we are in some sense decoupling the other non-Abelian gauge groups. 
Another limitation on $E_8$ is that given an $E_8$ symmetry on a single divisor it is still possible to enhance it further over subloci \cite{Morrison:1996pp}. Specifically, according to Kodaira's classification the discriminant of the elliptic fibration vanishes to order 10 over an $E_8$ singularity, but it is not difficult to construct geometries where it vanishes to a higher order, say 12, over subloci. However over such loci which enhance beyond $E_8$ one finds an infinite tower of massless degrees of freedom, the excitations of tensionless strings \cite{Witten:1995gx,Ganor:1996mu,Seiberg:1996vs,Witten:1996bn,Aspinwall:1997ye,Candelas:2000nc}. In F-theory these arise because in the singular limit some 4-cycle contracts to zero size over the loci extending $E_8$ and M5-branes wrapping this 4-cycle lead to the aforementioned strings.\footnote{In the Heterotic dual they arise from small instantons or in the M-theory picture from M2-branes stretching between M5-branes that are coinciding with the $E_8$ branes \cite{Witten:1995gx,Ganor:1996mu}.} If, for phenomenological reasons, we insist on the absence of these massless states then such extensions of $E_8$ are forbidden. 

Another clear limitation of the role of $E_8$ is that the group over the divisor is $SU(5)$ which is not in the exceptional branch of the Lie groups, and indeed is part of the infinite $SU(n)$ branch. Therefore by breaking a high enough $SU(n)$ down to $SU(5)$ one expects an, at least group theoretically, infinite number of matter spectra and charges to be realisable in F-theory.\footnote{There are mild constraints that make this finite due to tadpole cancellation but certainly the rank of the original $SU(n)$ can be much larger than that of $E_8$.} However we can appeal to phenomenological constraints once more to rule the possibility of the GUT group originating from a Higgsed down symmetry group which is in one of the infinite branches (the classical groups). This is because the Yukawa coupling for the top quark requires an exceptional symmetry enhancement over a co-dimension three point on the GUT brane \cite{Tatar:2006dc,Donagi:2008ca,Beasley:2008dc}, and such an exceptional structure can not come from any Higgsed classical group.

The requirement of a co-dimension three exceptional point rules out a Higgsed classical group possibility but neither does it imply that the matter spectrum and symmetries should come from a Higgsed exceptional group. The interplay between the symmetries at co-dimension three (Yukawas), co-dimension two (matter) and co-dimension one (gauge groups) is not well understood enough yet for us to be able to systematically restrict or classify the lower co-dimension data from the higher co-dimension data. The possibility that the existence of a point of exceptional symmetry implies that the whole theory on the GUT brane should be describable as a Higgsed down $E_8$ theory has been disproved by explicit examples: in \cite{Borchmann:2013hta} it was shown that some global models can not have an embedding inside a Higgsed $E_8$ theory, following earlier hints that this is so \cite{Mayrhofer:2012zy,Braun:2013yti}. This raises the question of whether $E_8$ has any role to play at all in bounding and classifying the possible F-theory GUT models?

The primary aim of this paper is to define, classify and study a certain well motivated extension to the set of theories obtained by Higgsing down an $E_8$ theory which can account for the global models in \cite{Borchmann:2013hta} as well as others in the literature. This set of theories will extend but be closely tied to $E_8$, and therefore show that $E_8$ may still have a role to play in the systematics of F-theory GUTs. It could be that the set of theories we will construct form a complete classification of possible GUT models in F-theory which include an exceptional point, no infinite tower of massless states, and are generic in a sense that we will define more precisely later. We have though no completely convincing evidence for this, but at the least they form a step towards understanding if such a complete classification exists and if so what it is.\footnote{There are certainly some theories that are not included in our classification: those reached by Higgsing only a chiral singlet rather than a vector-like pair, which should correspond to geometric gluing modes \cite{Cecotti:2010bp,Donagi:2011jy,Donagi:2011dv,Anderson:2013rka,Collinucci:2014qfa,Collinucci:2014taa}.} 

In short, the extended set of theories that we consider are constructed as follows. Consider the decomposition of the adjoint of $E_8$ under the breaking $E_8 \rightarrow SU(5)_{GUT} \times SU(5)_{\perp} \rightarrow SU(5)_{GUT} \times U(1)^4$. The adjoint will lead to 20 GUT-singlet fields $\un_i$ which have different charges under the Abelian group. Group theoretically these span the (off-diagonal components of the) adjoint representation of $SU(5)_{\perp}$. Therefore a theory which comes from Higgsing $E_8$ to $SU(5)_{GUT}$ is described by some appropriate vacuum expectation value for the singlets $\un_i$. The singlet fields form oppositely charged conjugate pairs and in this work we will only consider backgrounds where each element in the pair has equal vev. The set of representations that can appear in any theory that comes from a Higgsed $E_8$ is therefore determined by Higgsing a number of these 10 pairs of singlets. Each Higgsing will break a $U(1)$ in $U(1)^4$ eventually leaving a theory with no symmetries further to $SU(5)_{GUT}$. Our proposal is to extend this set of theories by adding a further differently charged 15 pairs of singlet fields to the $SU(5)_{GUT} \times U(1)^4$ theory, ones which do not come from the adjoint of $E_8$, and then construct the set of theories that can be reached by Higgsing also these new singlets. The motivation for adding these 15 new pairs is described in detail in the next section, but in brief they are the set of fields required such that for any pair of $SU(5)_{GUT}$-charged fields there is an associated gauge invariant cubic coupling with some singlet field, of type $\un \f \fb$. 

The result is a classification of some set of theories or more precisely the charges of the representations that can appear in the theories under any symmetry group present. It is important to state that we do not construct the F-theory geometries associated to all these theories, rather only present a group theoretic analysis of their representations. In section \ref{sec:embed}, we will then compare this set of theories with the `experimental data' of actual $SU(5)$ F-theory geometries constructed in the literature. We find that of 30 $SU(5)$ fibrations studied, 3 could not be made flat or generic enough in a sense defined more precisely below, and of the remaining 27, only one could be embedded into a Higgsed $E_8$ theory but all could be embedded into our extended set of theories.

Following this, in section \ref{sec:pheno}, we study some phenomenological applications of the new set of theories we have classified, in particular with respect to a realisation of ${\mathbb Z}_2$ matter parity which can not arise in a Higgsed $E_8$ theory. In section \ref{sec:hetd} we study aspects of the heterotic duals of these theories. We summarise our results in section \ref{sec:summary}.

\section{Global F-theory Models and $E_8$}
\label{sec:globfe8}

We are interested in models of the type $SU(5)\times U(1)^n$, where $n$ ranges between $0$ and $4$. Later we will also incorporate discrete symmetries into the framework. A key role will be played by representations that arise from the decomposition of the adjoint representation of $E_8$ into $SU(5)\times U(1)^4$. It is convenient to consider the embedding of the group into $E_8$ through an intermediate embedding of $E_8 \supset SU(5)_{GUT} \times SU(5)_{\perp}$. The $SU(5)_{GUT}$ factor is the remaining non-Abelian group while the $U(1)^4$ part is embedded as the Cartan subgroup of $SU(5)_{\perp}$. Then the decomposition of the adjoint of $E_8$ under $E_8 \rightarrow SU(5)_{GUT}\times SU(5)_{\perp}$ yields
\be
\bf{248} \rightarrow \left(\bf{24},\bf{1}\right)\op\left(\bf{1},\bf{24}\right)\op\left(\te,\f\right)\op\left(\fb,\te\right)\op\left(\teb,\fb\right)\op\left(\f,\teb\right) \;.
\ee
Therefore the GUT $\te$-multiplets are in the fundamental representation of $SU(5)_{\perp}$ and the GUT 5-multiplets are in the anti-symmetric of $SU(5)_{\perp}$. An embedding of a $U(1)$ into $SU(5)_{\perp}$ is specified by 5 parameters $a_i$ which determine its embedding into the Cartan $S\left[U(1)^5\right]$ and therefore should satisfy a tracelessness constraint $\sum_i a_i=0$. Our  notation is to write a particular $U(1)$ embedding as
\begin{equation}
U(1)_A = \sum_{i=1}^5 a^A_i t^i \;, \label{u1}
\end{equation}
where the $t^i$ are introduced to determine the $U(1)$ charges of the states as follows. With this embedding, from the adjoint of $E_8$ we find the following representations of $SU(5)_{GUT}$ with $U(1)$ charges labeled by $t_i$,
\begin{eqnarray}
\te_i: t_i \;, \;\;\;
\fb_{ij}: t_i+t_j \;, \;\;\;
\un_{ij}: t_i-t_j \;, \label{u1charges}
\end{eqnarray}
where for the $\fb$s and $\un$s we have that $i \neq j$. Here the $t_i$ correspond to the $U(1)$ charges of the representations in the sense that for a given $U(1)$, specified by \eqref{u1}, the charges are simply given by the contraction of the $t_i$ and $t^i$ using $t_i t^j = \delta_i^j$. Note that there are two types of gauge invariant operators which can be constructed from the fields in \eqref{u1charges}. There are operators whose charges $t_i$ sum to zero, for example $\f \, \te \, \te$ couplings, and operators, for example $\fb \, \fb \, \te$ couplings, whose $t_i$ sum to $t_1+t_2+t_3+t_4+t_5$.

So far we have discussed only group theory. Next we consider F-theory geometries that realise an $SU(5)$ gauge group on a divisor $W$ of the Calabi-Yau fourfold which projects down to a surface $S$ in the base of the fibration. The matter representations are taken to localise on curves in $S$. The cubic Yukawa couplings between $SU(5)$-charged fields occur at points where three such curves intersect. There are two types which we label as $E_6$ and $SO(12)$ and they correspond to couplings of the form $\fb\; \fb\; \te$ and $\f\; \te\; \te$ respectively. There are also cubic couplings of the form $\un\; \f\; \fb$, $\un\; \te\; \teb$ which occur at points where two matter curves intersect and a singlet curve intersects also the point, we label these $SU(7)$ and $E_6$ points respectively. In this paper we are interested in exploring the interaction between this class of F-theory models and the group $E_8$. 

It is useful to introduce some notation at this point. We define:
\begin{itemize}

\item A {\it network} as the data of the collection of $SU(5)$-charged matter curves on $S$ and their intersections. 

\item A {\it partially complete network} as a network where any pair of curves intersect each other at least once.

\item A {\it complete network} as a partially complete network where additionally any pair of $\f$ or $\te$ matter curves have a cubic coupling with a GUT singlet at some point.

\item A {\it flat network} as a network where for any point of intersection of two curves there is an associated cubic gauge-invariant coupling.

\end{itemize}

These definitions map to specific geometric properties of F-theory fibrations. A partially complete network maps to geometries where the base of the fibration is sufficiently generic, since on a non-generic base it may be that certain matter curves happen to not intersect. The data of the matter curves and interactions is then captured completely by the elliptic fibre equation. The notation of a flat network comes about because in F-theory constructions with Abelian symmetries, we find that if there is an intersection point which does not satisfy this criterion the point corresponds to a non-minimal singularity. Such co-dimension three non-minimal singularities were first studied in \cite{Candelas:2000nc} and were shown to be resolved into a non-flat fibration. There are therefore tensionless strings associated to them, which as discussed could signal going beyond $E_8$, and which we would like to avoid. Finally the difference between a complete network and a partially complete network is that there can exist partially complete networks where a pair of $\f$ matter curves only have a coupling with a $\te$ matter curve but no GUT singlet coupling. This would require specific geometric equations for the curves, examples of which can be found in \cite{Lawrie:2014uya}. Generally we will be interested in the relation between $E_8$ and complete flat networks.\footnote{We expect that for less generic configurations the relation with $E_8$ becomes more complicated. For example, one could consider a network which splits into two factors that do not share any intersections, and that the point of $E_6$ enhancement lies in only one factor. Then it is not clear why the other factor in the network of curves should be tied to the exceptional groups at all.}

Let us return now to the group theory analysis with the charges for the curves given in (\ref{u1charges}). Then a natural question is: can these charges form a complete flat network as defined above? It is easy to see that with regards to $E_6$ and $SO(12)$ points the charges are such that there is always an appropriate cubic combination of curves for each pair that is gauge invariant. However, this is not the case for $SU(7)$ points, i.e. not every pair of $\f$s has an appropriately charged singlet that could make a gauge neutral operator with it. Therefore the adjoint of $E_8$ does not have enough matter to form a complete flat network. It is worth noting for later that the lack of singlets in the adjoint of $E_8$ occurs only for $SU(5)$ as a GUT group. For $SO(10)$ or larger GUT groups the Abelian charges are such that the matter curves and singlets can always form a complete flat network. 

The natural conclusion one can reach from these considerations is that F-theory fibrations which form complete flat networks can have more singlet fields than those coming from a single adjoint representation of $E_8$. This observation leads to a natural extension of the spectrum of fields coming from the adjoint of $E_8$ as follows: For each pair of $\f$ matter curves we impose that there should be a singlet field such that there is a gauge invariant $\un\; \f\; \fb$ coupling. This extends the 10 singlets in (\ref{u1charges}) by a further 15 singlets. Starting from this extended set of fields we can now consider Higgsing down the number of $U(1)$s using not only the $E_8$ singlets but also the newly added ones. The resulting set of theories with a smaller Abelian sector will include the theories coming from a single adjoint Higgsing of $E_8$ only as a subset. They will form a set of theories that have charges which allow for a complete flat network which are based on $E_8$ but extend it. 

In the next section we will give an example geometry of this Higgsing process of non-$E_8$ singlets. In section \ref{HiggsBeyondE8} we will then construct the full set of resulting theories coming from Higgsing down beyond $E_8$. In section \ref{sec:embed} we will then proceed to compare this new set of theories with explicit F-theory geometries.

\subsection{An Example of Higgsing Beyond $E_8$}
\label{ExaBeyondE8}

In this section we present a geometric example which embodies the main ideas of the paper. We will consider the case of Higgsing $E_8$ down to $SU(5)\times U(1)$ with a Higgs bundle embedded in $S\left[U(3)\times U(2)\right] \supset SU(5)_{\perp}$. The matter states coming from the adjoint of $E_8$ under this decomposition are
\be
\te^1_{-2}\;,\; \te^2_{3}\;,\; \f^1_{-6} \;,\; \f^2_{4} \;,\; \f^3_{-1} \;,\; \un^1_{5} \;,
\ee
where the subscript denotes their charge under the $U(1)$. Note that there is a singlet of charge $10$ missing from the spectrum, which we denote, $\un^2_{10}$, that would be needed for this to form a complete network since there is no possible $\un^2_{10} \f^1_{-6}\fb^2_{-4}$ coupling. 

In \cite{Mayrhofer:2012zy} a global F-theory fibration based on this Higgsing of $E_8$ was constructed. The idea was to take the spectral cover description of this Higgsing process, which amounts to a 3-2 factorisation of the spectral cover \cite{Marsano:2009wr}, and restrict the fibration in Tate form such that the Tate coefficients match those of the spectral cover. In \cite{Mayrhofer:2012zy} two formulations of this construction were given, the first in Tate form, while the second was as a fibration in $P_{[1,1,2]}$. The Tate form of the fibration begins from the fibre equation
\be
y^2 = x^3  + a_1 x y z + a_2 x^2 z^2 + a_3 y z^3 + a_4 x z^4 + a_6 z^6 \;,
\ee
and restricts the coefficients to be of the form
\bea
a_1 &=& e_2 d_3 \;, \nonumber \\
a_2 &=& \left(e_2 d_2 + \alpha \delta d_3\right) w \;, \nonumber \\
a_3 &=& \left(\alpha \delta d_2 + \alpha \beta d_3 - e_2 \delta \gamma\right) w^2 \;, \nonumber \\
a_4 &=& \left(\alpha \beta d_2 + \beta e_2 \gamma - \alpha \delta^2 \gamma\right) w^3 \;, \nonumber \\
a_6 &=& \alpha \beta^2 \gamma w^5 \;.
\eea
Here $\alpha,\beta,\gamma,\delta,e_2,d_3$ are functions of the base coordinates, and $w=0$ is the $SU(5)$ divisor.
First let us map this into a fibration in $P_{[1,1,2]}$. We start from the form given in \cite{Morrison:2012ei} for the general single $U(1)$ fibration
\be
P_{[1,1,2]} = w^2 + b_0 u^2 w + b_1 u v w + b_2 v^2 w - u \left(c_0 u^3 + c_1 u^2 v + c_2 u v^2 + c_3 v^3 \right) \;, \label{p112fib}
\ee
and restrict the coefficients as follows \cite{Mayrhofer:2012zy}
\bea
\label{eq:coeff-3-2-g}
b_0 =    -w d_3 \alpha \;\;,  \;
b_1 =    -e_2 d_3  \;\; , \;
b_2 =    \delta \;\;,\;
c_0 =  w^3\,\alpha\,\gamma \;\;,  \;
c_1 =  w^2\,(d_2 \alpha + e_2 \gamma) \;\;,  \;
c_2 =   w\,e_2\, d_2 \;\;,  \;
c_3 =   w\,\beta \;.\nonumber
\eea
The $SU(5)$ singularity is over $w=0$ and the resulting matter curves are \cite{Marsano:2009wr,Mayrhofer:2012zy}
\bea
\te^1_{-2}&:& w=d_3=0 \;\;, \;\te^2_{3}:  w=e_2=0 \;, \nonumber \\
\f^1_{-6} &:& w=\delta=0, \qquad \f^2_{4} \;:\; w=\beta d_3 + d_2 \delta=0, \nonumber\\
\f^3_{-1} &:& w=\alpha^2 c_2 d_2^2 + \alpha^3 \beta d_3^2 + \alpha^3 d_2 d_3 \delta - 2 \alpha c_2^2 d_2 \gamma - 
  \alpha^2 c_2 d_3 \delta \gamma + c_2^3 \gamma^2 = 0.
\eea
This model has two singlet fields, with the second one being precisely the one lying outside of $E_8$. This singlet $\un^2_{10}$ is localised on the curve $\beta=\delta=0$ and so intersects the GUT brane at the point $w=\delta=\beta=0$ where also $\f^1_{-6}$ and $\f^2_{4}$ meet to form a cubic coupling. We therefore observe that in constructing the global F-theory geometry based on the Higgsed $E_8$ theory we automatically find the extra singlet required to make the complete network over a generic base. 

Note that this fibration has a non-minimal singularity at $\alpha=e_2=0$. Further, in \cite{Mayrhofer:2012zy} a resolution of this fibration was presented but there remained a singularity over $\alpha=\gamma=0$. These two issues can both be bypassed by setting $\alpha$ to a non-vanishing constant \cite{Mayrhofer:2012zy,Borchmann:2013hta}. 

So far we have seen a realisation of the first part of our extension of $E_8$: the inclusion of new singlets. The second part is the Higgsing of these singlets to reach new theories. We can present a geometric realisation of this by deforming the fibration in a way which corresponds to Higgsing the $\un^2_{10}$ singlet. This particular deformation has been recently studied in \cite{Braun:2014oya,Morrison:2014era,Garcia-Etxebarria:2014qua,Mayrhofer:2014haa,Mayrhofer:2014laa}. It amounts to adding a term $c_4 v^4$ to (\ref{p112fib}). We just need to modify it slightly by a factor of $w$ in order to account for the additional $SU(5)$  so we write $c_4 = c_{4,1} w$.

After the deformation one finds that the two matter curves $\f^1_{-6}$ and $\f^2_{4}$ recombine into a single matter curve with equation
\be
\tilde{\f}^1 : \delta \left( \beta d_3 + d_2 \delta \right) + e_2 c_{4,1} d_3^2 = w = 0\;. 
\ee
The other matter curves remain unchanged. The Higgsing deformation performed breaks the $U(1)$ to a ${\mathbb Z}_2$. The resulting charges of the matter curves under this ${\mathbb Z}_2$ are
\be
\te^1_{0}\;,\; \te^2_{1}\;,\; \tilde{\f}^1_{0} \;,\; \f^3_{1} \;,\; \un^1_{1} \;,
\ee
This model does not lie in a possible Higgsing of $E_8$ and we therefore reach a new model precisely through the process described in the previous sections. As an aside, note that this is the first example of such a ${\mathbb Z}_2$ model with two $\te$-matter curves. 

It is interesting to map the Higgsing back into the factorised Tate form. The Higgsing then amounts to a deformation
\bea
a_4 &\rightarrow& a_4 - c_{4,1} \alpha \left(\alpha d_3^2 + 4 \gamma w\right)  w^3\;, \nonumber \\
a_6 &\rightarrow& a_6 + c_{4,1} \left(-\alpha d_2 + e_2 \gamma\right)^2 w^5 \;.
\eea
An important point is that the Higgsing involves the next-to-leading order in $w$ coefficient of $a_4$, or in the usual notation $ a_{4,4} \rightarrow a_{4,4} + 4 c_{4,1} \alpha \gamma$. This modification ensures the correct discriminant enhancement at the coupling point $\un^1_{1}\; \tilde{\f}^1_{0}\; \fb^3_{1}$.\footnote{Indeed as pointed out in \cite{Borchmann:2013hta}, generally the $\un \f \fb$ couplings depend on the higher order terms in $w$ in the Tate coefficients. Note that this is not inconsistent with the fact that the point of coupling can be determined purely from the leading order terms of coefficients since it corresponds to the intersection of two $\f$-matter curves. Indeed the global aspect of a $U(1)$, or section, ensures that the sub-leading parts are such that there is an appropriate discriminant enhancement at the intersection of two $\f$-matter curves.} Note that by contrast Higgsing the singlets from the adjoint of $E_8$, in this case $\un^1_{5}$, can be done by deforming just the leading order coefficients in the Tate fibration. An example where this is clear is the Higgsing of the factorised Tate 4-1 model \cite{Borchmann:2013hta}. In the case of a model based on $S\left[U(4)\times U(1)\right] \supset SU(5)_{\perp}$ one finds that there are only 2 $\f$-matter curves and the singlet at their intersection has an embedding in the adjoint of $E_8$. A particular restriction of this fibration gives the $U(1)$-restricted Tate model \cite{Grimm:2010ez} which simply amounts to setting $a_6=0$ in the generic Tate polynomial. Then Higgsing the singlet can be done by taking $a_{6,5} \neq 0$ and leaving the other coefficients unchanged, a deformation which only affects therefore the leading order (in $w$) behaviour of the Tate coefficients.

The Higgsing away from $E_8$ studied in this section forms a small branching in the full classification of such possible Higgsing to which we now turn.

\subsection{Classifying Higgsing Beyond $E_8$}
\label{HiggsBeyondE8}

Recall the $E_8$ singlets are defined through their charges as in (\ref{u1charges}). Similarly we define the 15 additional beyond $E_8$ singlets in terms of the $t_i$ as
\begin{equation}
 \mathbf{1}_{ijkl}:~t_i+t_j-t_k-t_l \;,
\end{equation}
where $i\neq j \neq k \neq l$. This set of states ensures that there exists a gauge-invariant coupling for every pair of $\f$ states.

The set of theories we wish to study are defined as starting from the maximal decomposition of $E_8$ to $SU(5)_{GUT}\times U(1)^4$ and then Higgsing all the possible combinations of GUT singlets in the theory. The gauge group that remains $G$ will be the commuting subgroup of $SU(5)_{GUT}\times U(1)^4$ with all the Higgsed singlets and the matter representations of the theory will be the representations of $G$ that descend from the adjoint of $E_8$ plus the additional singlets. We denote the set of Higgsed singlets by $\un_{\alpha}$, with $\alpha$ ranging over the different singlets $\alpha=1,...,N$. The charges of a given singlet $\un_{\alpha}$ under $U(1)^5$, i.e. before imposing the tracelessness constraint, are denoted $Q_i\left(\un_{\alpha}\right)$. To work out the group $G$ and its representations for a given set of Higgsed singlets we introduce a $5 \times N$ matrix ${\cal M}$ of the Higgsed singlets charges
\be
\mathcal{M}_{i\alpha}=Q_i\left(\mathbf{1}_{\alpha}\right) \;.
\ee
The tracelessness constraint is implemented in this framework by including in the Higgs matrix ${\cal M}$ a singlet with charges $\left(1,1,1,1,1\right)$, which breaks the $U(1)^5$ to $S\left[U(1)^5\right]$. To proceed, following the methodology of \cite{Petersen:2009ip}, we change the basis of $U(1)$s such that each Higgsed singlet is charged under only one $U(1)$ combination. This is the Smith Form ${\cal D}$ associated to ${\cal M}$ which is reached by acting with two unimodular matrices ${\cal K} \in \text{SL}_N(\mathbb{Z})$ and ${\cal J} \in \text{SL}_5\left(\mathbb{Z}\right)$ 
\be
{\cal K} {\cal M} {\cal J} = {\cal D} = \text{diag}(d_1,\dots,d_r,0,\dots,0) \;,
\ee
where the integer entry $d_{i-1}$ divides $d_{i}$ for all $i=2,\dots,r=\text{Rank}(\mathcal{D})$. The matrix ${\cal J}$ determines the appropriate change in basis of $U(1)$s such that the charge vector $Q_i$ of any state transforms to $Q'_i = \left( Q {\cal J}\right)_{i}$. In particular the integer charges under the unbroken gauge group $G$ are therefore given by $Q'_i$ for $i=r+1,...,5$.

Note that some of the $U(1)$s may be broken to a remnant discrete symmetry if their charges are not unitary. Since the Smith form is reached by a unimodular transformation the remnant discrete gauge group is simply 
\be
G_{\mathrm{Discrete}} = \mathbb{Z}_{d_1}\times\dots\times\mathbb{Z}_{d_r} \;.
\ee
Note that some of $G_{\mathrm{Discrete}}$ may be part of the $\mathbb{Z}_5$ centre of $SU(5)$ in which case the true discrete subgroup would be $G_{\mathrm{Discrete}}/\mathbb{Z}_5$.

\begin{sidewaysfigure}[p]
  \centering
  \includegraphics[width=600pt]{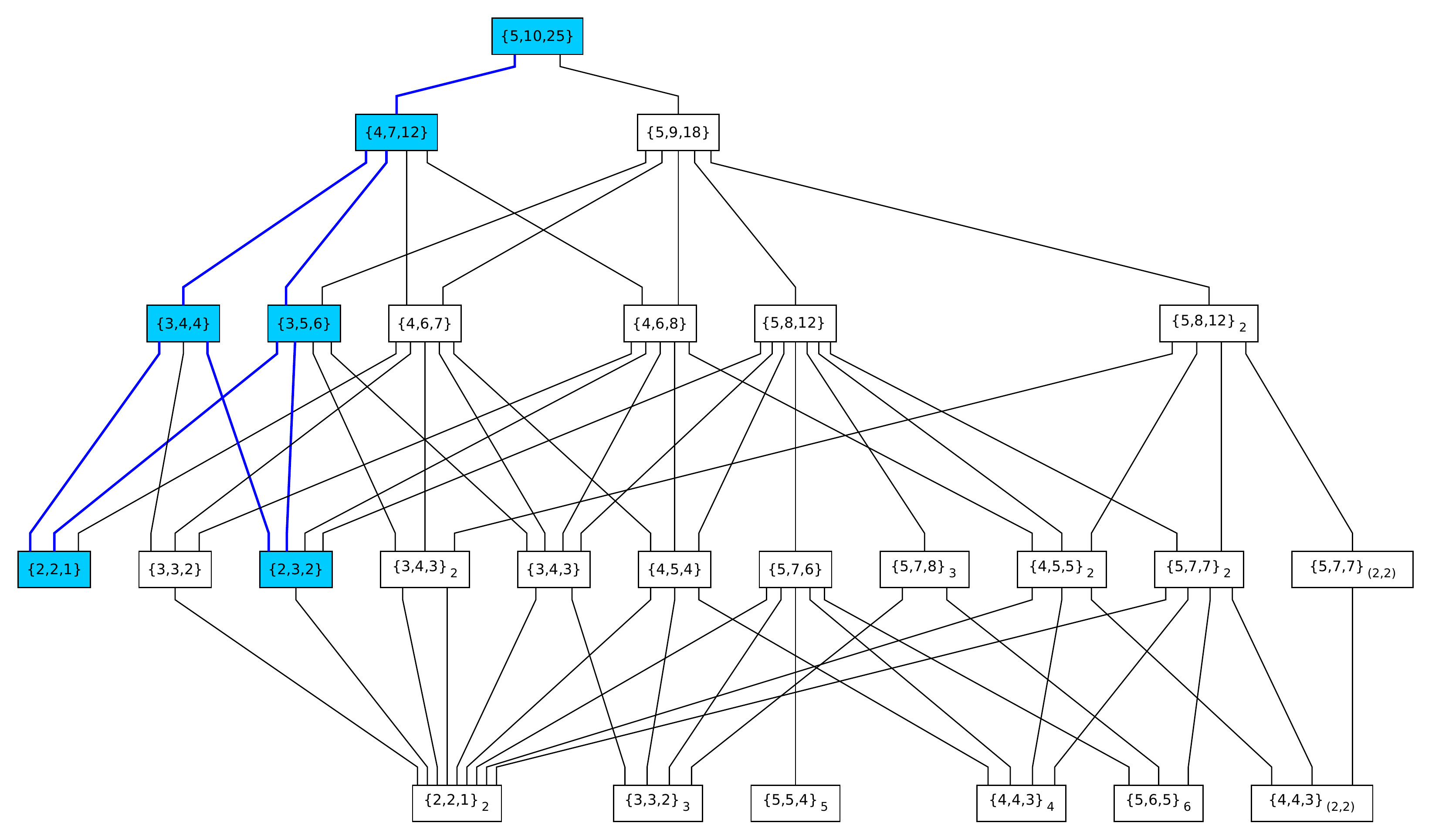}
  \caption{The set of theories that can be reached by Higgsing down from $SU(5)\times U(1)^4$ using the GUT singlets. Each rectangle denotes a spectrum of fields, with the numbers giving the number of differently charged $\te$, $\f$ and $\un$ representations, and the subscript denoting the order of any discrete group present. The spectra of all the models are given in tables 2.1 and 2.2. The paths connecting the models denote Higgsing of GUT singlets. The blue nodes and paths correspond to the set of theories reached from adjoint Higgsing of $E_8$, though the spectrum of singlets in these models is extended beyond $E_8$. The decreasing levels denote a decreasing number of $U(1)$s.}
	\label{modelpaths}
\end{sidewaysfigure}

The above procedure of calculating $G$ from a given set of Higgsed singlets must be performed for all the possible combinations of Higgsing the singlets. Since the initial theory has 25 singlets we have 25 possibilities for Higgsing 1 singlet, 300 possibilities for Higgsing 2 singlets, 2300 possibilities for Higgsing 3 singlets and 12650 possibilities for Higgsing 4 singlets. However there is significant redundancy in this counting and the final set of physically independent possibilities is much smaller in magnitude. For example it is clear that for the case of 1 singlet there are only 2 physically distinct possibilities: Higgsing a $\un_{ij}$ or a $\un_{ijkl}$ state, with all such sets being equivalent up to relabeling. The similar analysis of redundancies can be feasibly performed for 2 singlets analytically, but is much harder for more singlets. We performed this analysis using computer scanning, with some analytic checks. The final result is shown in tables 2.1 and 2.2. The models are labeled by three numbers denoting the number of differently charged $\te$, $\f$ and $\un$ representations respectively. The number of physically distinct models with 3, 2, 1 and 0 $U(1)$s is given by 2, 6, 11, 6 respectively. The paths which can be taken to reach each of the models as a Higgsing process are shown in figure \ref{modelpaths}. 

\begin{sidewaystable}
	\centering
	\begin{tabular}{c||c|c|c|c|c || c|c|c|c|c|c|c|c|c}
		\tiny Model & $\mathbf{10}_1$ & $\mathbf{10}_2$ & $\mathbf{10}_3$ & $\mathbf{10}_4$ & $\mathbf{10}_5$ &

		 $\mathbf{\bar{5}}_1$ & $\mathbf{\bar{5}}_2$ & $\mathbf{\bar{5}}_3$ & $\mathbf{\bar{5}}_4$ & $\mathbf{\bar{5}}_5$ & $\mathbf{\bar{5}}_6$ & $\mathbf{\bar{5}}_7$ & $\mathbf{\bar{5}}_8$ & $\mathbf{\bar{5}}_9$ \\\hline\hline
		&\multicolumn{14}{c}{Three $U(1)$'s models}\\\hline\hline
		\tiny$\mathbf{\left\{ 4,7,12 \right\}}$ & \tiny$(-2,-1,-1)$ & \tiny$(1,0,0)$ & \tiny$(0,1,0)$ & \tiny$(0,0,1)$ & --- & \tiny$(-2,-1,0)$ & \tiny$(-2,0,-1)$ & \tiny$(-1,-1,-1)$ & \tiny$(1,1,0)$ & \tiny$(1,0,1)$ & \tiny$(0,1,1)$ & \tiny$(2,0,0)$ &---&---\\\hline
		 &\multicolumn{14}{l}{ \tiny$(0,1,-1)$, \tiny$(1,-1,0)$, \tiny$(1,0,-1)$,\tiny$(4,1,0)$, \tiny$(4,0,1)$, \tiny$(2,2,1)$, \tiny$(2,1,2)$, \tiny$(1,2,2)$, \tiny$(2,-1,-1)$, \tiny$(3,2,0)$, \tiny$(3,0,2)$, \tiny$(3,1,1)$} \\\hline
		 \tiny$\left\{ 5,9,18 \right\}$ & \tiny$(-2,-2,0)$ & \tiny$(1,0,0)$ & \tiny$(0,1,0)$ & \tiny$(0,0,1)$ & \tiny$(1,1,-1)$ & \tiny$(-2,-2,1)$ & \tiny$(-2,-1,0)$ & \tiny$(-1,-2,0)$ & \tiny$(1,1,0)$ & \tiny$(1,0,1)$ & \tiny$(0,1,1)$ & \tiny$(-1,-1,-1)$ & \tiny$(1,2,-1)$ & \tiny$(2,1,-1)$ \\\hline
		 &\multicolumn{14}{l}{ \tiny$(4,3,-2)$, \tiny$(4,2,-1)$, \tiny$(3,4,-2)$, \tiny$(3,3,-1)$, \tiny$(3,2,0)$, \tiny$(3,1,1)$, \tiny$(2,4,-1)$, \tiny$(2,3,0)$, \tiny$(2,2,1)$, \tiny$(2,1,2)$, \tiny$(2,0,-2)$, \tiny$(1,3,1)$, \tiny$(1,2,2)$, \tiny$(1,1,-2)$, \tiny$(1,0,-1)$, \tiny$(1,-1,0)$, \tiny$(0,2,-2)$, \tiny$(0,1,-1)$ }\\\hline\hline
		&\multicolumn{14}{c}{Two $U(1)$'s models}\\\hline\hline
		\tiny$\mathbf{\left\{ 3,4,4 \right\}}$ & \tiny$(-3,-1)$ & \tiny$(1,0)$ & \tiny$(0,1)$ & --- & --- & \tiny$(-3,0)$ & \tiny$(-2,-1)$ & \tiny$(1,1)$ & \tiny$(2,0)$ & --- & --- & --- & --- & --- \\\hline
		 &\multicolumn{14}{l}{ \tiny$(1,-1)$, \tiny$(3,2)$, \tiny$(4,1)$, \tiny$(5,0)$}\\\hline
		\tiny$\mathbf{\left\{ 3,5,6 \right\}}$ & \tiny$(-2,-2)$ & \tiny$(1,0)$ & \tiny$(0,1)$ & --- & --- & \tiny$(-2,-1)$ & \tiny$(-1,-2)$ & \tiny$(1,1)$ & \tiny$(2,0)$ & \tiny$(0,2)$ & ---& --- & --- & --- \\\hline
		 &\multicolumn{14}{l}{\tiny$(1,-1)$, \tiny$(1,4)$, \tiny$(2,-2)$, \tiny$(2,3)$, \tiny$(3,2)$, \tiny$(4,1)$}\\\hline
		 \tiny$\left\{ 4,6,7 \right\}$ & \tiny$(-1,2)$ & \tiny$(0,-4)$ & \tiny$(1,0)$ & \tiny$(0,1)$ & --- & \tiny$(-1,-2)$ & \tiny$(-1,3)$ & \tiny$(0,-3)$ & \tiny$(0,2)$ & \tiny$(1,-4)$ & \tiny$(1,1)$ & --- & --- & --- \\\hline
		 &\multicolumn{14}{l}{\tiny$(0,5)$, \tiny$(1,-6)$, \tiny$(1,-1)$, \tiny$(1,4)$, \tiny$(2,-7)$, \tiny$(2,-2)$, \tiny$(2,3)$} \\\hline
		 \tiny$\left\{ 4,6,8 \right\}$ & \tiny$(-2,-2)$ & \tiny$(0,1)$ & \tiny$(1,0)$ & \tiny$(3,3)$ & --- & \tiny$(-4,-4)$ & \tiny$(-2,-1)$ & \tiny$(-1,-2)$ & \tiny$(1,1)$ & \tiny$(3,4)$ & \tiny$(4,3)$ & --- & --- & --- \\\hline
		 &\multicolumn{14}{l}{\tiny$(1,-1)$, \tiny$(2,3)$, \tiny$(3,2)$, \tiny$(4,6)$, \tiny$(5,5)$, \tiny$(6,4)$, \tiny$(7,8)$, \tiny$(8,7)$}\\\hline
		 \tiny$\left\{ 5,8,12 \right\}$ & \tiny$(-4,6)$ & \tiny$(-1,1)$ & \tiny$(0,1)$ & \tiny$(2,-4)$ & \tiny$(3,-4)$ & \tiny$(-5,7)$ & \tiny$(-4,7)$ & \tiny$(-2,2)$ & \tiny$(-1,2)$ & \tiny$(1,-3)$ & \tiny$(2,-3)$ & \tiny$(3,-3)$ & \tiny$(5,-8)$ & --- \\\hline
		 &\multicolumn{14}{l}{\tiny$(1,0)$, \tiny$(2,-5)$, \tiny$(2,0)$, \tiny$(3,-5)$, \tiny$(4,-5)$, \tiny$(5,-10)$, \tiny$(5,-5)$, \tiny$(6,-10)$, \tiny$(7,-10)$, \tiny$(8,-10)$, \tiny$(9,-15)$, \tiny$(10,-15)$} \\\hline
		 \tiny$\left\{ 5,8,12 \right\}_2$ & \tiny$(-2,-2)_0$ & \tiny$(1,0)_0$ & \tiny$(0,1)_0$ & \tiny$(1,0)_1$ & \tiny$(0,1)_1$ & \tiny$(-2,-1)_0$ & \tiny$(-1,-2)_0$ & \tiny$(1,1)_0$ & \tiny$(-2,-1)_1$ & \tiny$(-1,-2)_1$ & \tiny$(1,1)_1$ & \tiny$(2,0)_1$ & \tiny$(0,2)_1$ & --- \\\hline
		 &\multicolumn{14}{l}{\tiny$(1,-1)_0$, \tiny$(1,4)_0$, \tiny$(2,-2)_0$, \tiny$(2,3)_0$, \tiny$(3,2)_0$, \tiny$(4,1)_0$, \tiny$(0,0)_1$, \tiny$(1,-1)_1$, \tiny$(1,4)_1$, \tiny$(2,3)_1$, \tiny$(3,2)_1$, \tiny$(4,1)_1$} \\\hline\hline

		&\multicolumn{14}{c}{One $U(1)$ models}\\\hline\hline

		\tiny$\mathbf{\left\{ 2,2,1 \right\}}$ & \tiny$-4$ & \tiny$1$ & --- & --- & --- & \tiny$-3$ & \tiny$2$ & --- & --- & --- & --- & --- & --- & --- \\\hline
		&\multicolumn{14}{l}{\tiny$5$}\\\hline
		\tiny$\mathbf{\left\{ 2,3,2 \right\}}$ & \tiny$-3$ & \tiny$2$ & --- & --- & --- & \tiny$-6$ & \tiny$-1$ & \tiny$4$ & --- & --- & --- & --- & --- & ---\\\hline
		&\multicolumn{14}{l}{\tiny$5$, \tiny$10$}\\\hline
		 \tiny$\left\{ 3,3,2 \right\}$ & \tiny$-1$ & \tiny$0$ & \tiny$1$ & --- & --- & \tiny$-1$ & \tiny$0$ & \tiny$1$ & --- & --- & --- & --- & --- & ---\\\hline
		&\multicolumn{14}{l}{\tiny$1$, \tiny$2$}\\\hline
	\end{tabular}
	\label{u1theories1}
	\caption{First part of the summary of the $SU(5)$-charged spectra for the models reached by Higgsing $SU(5) \times U(1)^4$. The numbers indicate charges under the $U(1)$s present, with subscripts indicating a discrete charge. The second row for each model lists the GUT singlets present. Models in bold are models accessible by Higgsing only $E_8$ singlets and therefore have charged matter spectra arising from the adjoint of $E_8$ but with additional singlets.}
\end{sidewaystable}

\begin{sidewaystable}
	\centering
	\begin{tabular}{c||c|c|c|c|c || c|c|c|c|c|c|c|c|c}
		\tiny Model & $\mathbf{10}_1$ & $\mathbf{10}_2$ & $\mathbf{10}_3$ & $\mathbf{10}_4$ & $\mathbf{10}_5$ &

		 $\mathbf{\bar{5}}_1$ & $\mathbf{\bar{5}}_2$ & $\mathbf{\bar{5}}_3$ & $\mathbf{\bar{5}}_4$ & $\mathbf{\bar{5}}_5$ & $\mathbf{\bar{5}}_6$ & $\mathbf{\bar{5}}_7$ & $\mathbf{\bar{5}}_8$ & $\mathbf{\bar{5}}_9$ \\\hline\hline
		&\multicolumn{14}{c}{One $U(1)$ models (Continued)}\\\hline\hline
		 \tiny$\left\{ 3,4,3 \right\}$ & \tiny$-4$ & \tiny$1$ & \tiny$6$ & --- & --- & \tiny$-8$ & \tiny$-3$ & \tiny$2$ & \tiny$7$ & --- & --- & --- & --- & ---\\\hline
		&\multicolumn{14}{l}{\tiny$5$, \tiny$10$, \tiny$15$}\\\hline
		 \tiny$\left\{ 4,5,4 \right\}$ & \tiny$-8$ & \tiny$-3$ & \tiny$2$ & \tiny$7$ & --- & \tiny$-11$ & \tiny$-6$ & \tiny$-1$ & \tiny$4$ & \tiny$9$ & --- & --- & --- & ---\\\hline
		&\multicolumn{14}{l}{\tiny$5$, \tiny$10$, \tiny$15$, \tiny$20$}\\\hline
		 \tiny$\left\{ 5,7,6 \right\}$ & \tiny$-2$ & \tiny$-1$ & \tiny$0$ & \tiny$1$ & \tiny$2$ & \tiny$-3$ & \tiny$-2$ & \tiny$-1$ & \tiny$0$ & \tiny$1$ & \tiny$2$ & \tiny$3$ & --- & ---\\\hline
		&\multicolumn{14}{l}{\tiny$1$, \tiny$2$, \tiny$3$, \tiny$4$, \tiny$5$, \tiny$6$}\\\hline
		 \tiny$\left\{ 3,4,3 \right\}_2$ & \tiny$-4_0$ & \tiny$1_0$ & \tiny$1_1$ & --- & --- & \tiny$-3_0$ & \tiny$-3_1$ & \tiny$2_0$ & \tiny$2_1$ & --- & --- & --- & --- & ---\\\hline
		&\multicolumn{14}{l}{\tiny$0_1$, \tiny$5_0$, \tiny$5_1$}\\\hline
		 \tiny$\left\{ 4,5,5 \right\}_2$ & \tiny$-3_0$ & \tiny$-3_1$ & \tiny$2_0$ & \tiny$2_1$ & --- & \tiny$-6_1$ & \tiny$-1_0$ & \tiny$-1_1$ & \tiny$4_0$ & \tiny$4_1$ & --- & --- & --- & ---\\\hline
		&\multicolumn{14}{l}{\tiny$0_1$, \tiny$5_0$, \tiny$5_1$, \tiny$10_0$, \tiny$10_1$}\\\hline
		 \tiny$\left\{ 5,7,7 \right\}_2$ & \tiny$-4_0$ & \tiny$-4_1$ & \tiny$1_0$ & \tiny$1_1$ & \tiny$6_0$ & \tiny$-8_1$ & \tiny$-3_0$ & \tiny$-3_1$ & \tiny$2_0$ & \tiny$2_1$ & \tiny$7_0$ & \tiny$7_1$ & --- & ---\\\hline
		&\multicolumn{14}{l}{\tiny$0_1$, \tiny$5_0$, \tiny$5_1$, \tiny$10_1$, \tiny$10_0$, \tiny$15_0$, \tiny$15_1$}\\\hline
		 \tiny$\left\{ 5,7,8 \right\}_3$ & \tiny$-3_1$ & \tiny$-3_2$ & \tiny$2_0$ & \tiny$2_1$ & \tiny$2_2$ & \tiny$-6_0$ & \tiny$-1_0$ & \tiny$-1_1$ & \tiny$-1_2$ & \tiny$4_0$ & \tiny$4_1$ & \tiny$4_2$ & --- & ---\\\hline
		&\multicolumn{14}{l}{\tiny$0_1$, \tiny$0_2$, \tiny$5_0$, \tiny$5_1$, \tiny$5_2$, \tiny$10_0$, \tiny$10_1$, \tiny$10_2$}\\\hline
		 \tiny$\left\{ 5,7,7 \right\}_{2,2}$ & \tiny$-4_{0,0}$ & \tiny$1_{0,0}$ & \tiny$1_{0,1}$ & \tiny$1_{1,0}$ & \tiny$1_{1,1}$ & \tiny$-3_{0,0}$ & \tiny$-3_{0,1}$ & \tiny$-3_{1,0}$ & \tiny$-3_{1,1}$ & \tiny$2_{0,1}$ & \tiny$2_{1,0}$ & \tiny$2_{1,1}$ & --- & ---\\\hline
		&\multicolumn{14}{l}{\tiny$0_{0,1}$, \tiny$0_{1,0}$, \tiny$0_{1,1}$, \tiny$5_{0,0}$, \tiny$5_{0,1}$, \tiny$5_{1,0}$, \tiny$5_{1,1}$}\\\hline\hline
		&\multicolumn{14}{c}{Zero $U(1)$ models}\\\hline\hline
		 \tiny$\left\{2,2,1\right\}_2$ & \tiny$0$ & \tiny$1$ & --- & --- & --- & \tiny$0$ & \tiny$1$ & --- & --- & --- & --- & --- & --- & ---\\\hline
		&\multicolumn{14}{l}{\tiny$1$}\\\hline
		 \tiny$\left\{3,3,2\right\}_3$ & \tiny$0$ & \tiny$1$ & \tiny$2$ & --- & --- & \tiny$0$ & \tiny$1$ & \tiny$2$ & --- & --- & --- & --- & --- & ---\\\hline
		&\multicolumn{14}{l}{\tiny$1$, \tiny$2$}\\\hline
		\tiny$\left\{4,4,3\right\}_{(2,2)}$ & \tiny$(0,0)$ & \tiny$(0,1)$ & \tiny$(1,0)$ & \tiny$(1,1)$ & --- & \tiny$(0,0)$ & \tiny$(0,1)$ & \tiny$(1,0)$ & \tiny$(1,1)$ & --- & --- & --- & --- & ---\\\hline
		&\multicolumn{14}{l}{\tiny$(0,1)$, \tiny$(1,0)$, \tiny$(1,1)$}\\\hline		
		\tiny$\left\{4,4,3\right\}_{4}$ & \tiny$0$ & \tiny$1$ & \tiny$2$ & \tiny$3$ & --- & \tiny$0$ & \tiny$1$ & \tiny$2$ & \tiny$3$ & --- & --- & --- & --- & ---\\\hline
		&\multicolumn{14}{l}{\tiny$1$, \tiny$2$, \tiny$3$}\\\hline		
		 \tiny$\left\{5,5,4\right\}_5$ & \tiny$0$ & \tiny$1$ & \tiny$2$ & \tiny$3$ & \tiny$4$ & \tiny$0$ & \tiny$1$ & \tiny$2$ & \tiny$3$ & \tiny$4$ & --- & --- & --- & ---\\\hline
		&\multicolumn{14}{l}{\tiny$1$, \tiny$2$, \tiny$3$, \tiny$4$}\\\hline
		 \tiny$\left\{5,6,5\right\}_6$ & \tiny$0$ & \tiny$1$ & \tiny$2$ & \tiny$4$ & \tiny$5$ & \tiny$0$ & \tiny$1$ & \tiny$2$ & \tiny$3$ & \tiny$4$ & \tiny$5$ & --- & --- & ---\\\hline
		&\multicolumn{14}{l}{\tiny$1$, \tiny$2$, \tiny$3$, \tiny$4$, \tiny$5$}\\\hline

	\end{tabular}
	\label{u1theories2}
	\caption{Second part of the summary of the $SU(5)$-charged spectra for the models reached by Higgsing $SU(5) \times U(1)^4$. The numbers indicate charges under the $U(1)$s present, with subscripts indicating a discrete charge. For the pure discrete symmetry models the discrete charges are not sub-scripted. The second row for each model lists the GUT singlets present. Models in bold are models accessible by Higgsing only $E_8$ singlets and therefore have charged matter spectra arising from the adjoint of $E_8$ but with additional singlets.}
\end{sidewaystable}

The final set of models with no $U(1)$ symmetry are differentiated purely by their discrete symmetries. Indeed it is interesting to note that models with discrete symmetries lie outside the Higgsed $E_8$ subset, marked in bold in the tables, so discrete symmetries are only induced by Higgsing non-$E_8$ singlets. However there is a set of discrete symmetries which are not captured by our analysis. From the perspective of a Higgsed $E_8$ theory these arise from symmetries lying in $SU(5)_{\perp}$ which are not embedded in its Cartan subgroup. They occur when the Higgs vev is restricted beyond just which components are non-vanishing but there are relations between the non-vanishing components.\footnote{In terms of the Spectral Cover approach such symmetries occur when the Galois group of the roots of the spectral cover is not a product of permutation groups (dictated by the $U(1)$ factorisation) but sub-groups of them. Or using earlier terminology when the monodromy group is not the full permutation group. See \cite{Heckman:2009mn,Marsano:2009gv,Dudas:2009hu,Antoniadis:2013joa,Karozas:2014aha} for studies of this.} These non-generic Higgs backgrounds can lead to models with discrete symmetries that come from a Higgsed $E_8$. We have not attempted to implement these in our classification because it is not clear what the prescription should be to extend these beyond the Higgsed $E_8$ picture. It would be interesting to understand such symmetries better from a global perspective and thereby gain some intuition as to how they may be implemented beyond a Higgsed $E_8$.

\subsection{Embedding Known Models}
\label{sec:embed}

In the previous section we derived a class of theories that were specified by the symmetries and representations present. Although motivated from F-theory geometry the analysis was purely group theoretic. The correspondence between the theories described and F-theory geometry should be essentially that the massless gauge fields and matter modes in the geometry match those of the theories, and that the gauge invariant couplings between fields correspond to points on $S_{GUT}$ where matter curves intersect. However there are some differences to be expected which we should outline. 

Of course the $SU(5)_{GUT}$ factor should match the associated non-Abelian singularity in F-theory and its corresponding massless matter. The Abelian symmetries could correspond to massless Abelian symmetries in F-theory, which would be additional sections of the fibration. However in principle they could also correspond to massive Abelian symmetries. If the symmetries are made massive through a Higgsing process, which geometrically is a complex-structure deformation, then we expect this to be the same as Higgsing a GUT singlet in our analysis and so the theory would just flow to a theory with less $U(1)$s in our classification. In type IIB string theory we also know of two other ways that a $U(1)$ could become massive which do not correspond to a Higgsing by an open-string mode. The first is through coupling to some background flux, and the second is through a geometric mass, see for example \cite{Jockers:2004yj,Grimm:2011dj,Grimm:2011tb}. In F-theory these are expected to uplift to backgrounds supporting G-flux, which can lead to a mass for a $U(1)$ as studied in detail for example in \cite{Cvetic:2012xn}, and to backgrounds which include a particular set of non-closed forms as studied in \cite{Grimm:2010ez,Grimm:2011tb}. Since we are considering geometries with no background flux the first potential mass term should be absent. The second mass term we expect to also be absent, at least in the geometric constructions that we will consider, though since it is not fully understood in an F-theory context one can not rule the possibility that it is present. One piece of evidence for the absence of a hidden massive $U(1)$ is that the discriminant splits into components which have differing charges under the massless Abelian sector. If there was a massive $U(1)$ then one would expect further factorisation, as occurs for example in the models where the $U(1)$ is Higgsed to a discrete symmetry \cite{Garcia-Etxebarria:2014qua,Mayrhofer:2014haa}. Further, there are no additional selection rules on Yukawa couplings in the geometries we will study beyond those of the $U(1)$s (and possibly discrete remnants from Higgsing). Finally, it was argued in \cite{Grimm:2010ez,Grimm:2011tb,Braun:2014nva} that one would expect such mass $U(1)$s would lead to geometries which do not allow for a CY resolution.\footnote{Note though that this was refined in \cite{Mayrhofer:2014laa} where it was shown that the non-CY element is present only if the discrete symmetry is present already in M-theory, while if it only emerges in the F-theory limit then it is possible to present a CY fibration.} Therefore in looking to how geometric models constructed are embedded in our classification we will take the most constrained embedding where the Abelian sector is completely massless. We should however, in the absence of a solid proof, keep in mind the possibility that a given geometric model can still be embedded inside one of our theories with a larger Abelian sector and then the charges of the matter fields would be under some subgroup of this large Abelian symmetry group which is the massless sector.

The matter representations should correspond to matter curves. In a given F-theory geometry we certainly should not expect all the possible representations in our theories to be present. An embedding of a geometric model in our classification should entail showing that the massless matter present in an F-theory model forms a subset of the representations in our model. The additional data of whether a representation corresponds to actual massless matter is purely geometric and matter curves can indeed be turned on and off by appropriate choices of fibrations. 

We now come to consider which class of F-theory geometries we might expect to be captured by our classification. Using the definitions in section \ref{sec:globfe8} it is most natural to study embeddings of F-theory geometries which form flat complete networks into our classification. The flatness criterion affects most of the models in the literature since they exhibit non-flat points. However, given a construction, the elliptic fibration can be restricted further to turn off these non-flat points by choosing the parameters of the fibration not generic but setting some to constants. Turning off the non-flat points in this way restricts the fibration further and may turn off some of the matter loci. It is this restricted fibration that we then attempt to embed into our set of theories. The criterion of a partially complete network simply amounts to considering a generic base for the fibration which we therefore assume in our embeddings. Finally, although most of the constructions in the literature form complete networks there are a few which only form partially complete networks. We will discuss these special cases below.

In table \ref{tab:embed} we give the possible embeddings of models in the literature into our classification. We considered 30 fibrations. 8 of these had an SU(5) charged matter spectrum which was embeddable in a Higgsed $E_8$ theory, but apart from one (the $4-1$ factorised Tate of \cite{Mayrhofer:2012zy}) all of them also had GUT singlets which were not embeddable in the adjoint of $E_8$. Therefore 29 models were in fact not embeddable in a Higgsed $E_8$ theory. One model could not be made flat over a generic base, and of the remaining 29, once they were constrained to be flat, 27 could be embedded into our classification. We present the analysis of restricting the fibrations to be flat in appendix \ref{app:nf}. In addition we did not list the four models constructed in \cite{Mayrhofer:2012zy} with more that one $U(1)$ which were based on a global extension of Higgsed $E_8$ theories (because a smooth resolution of them was not presented). There the results are known by construction: the charged matter spectrum can be embedded in a Higgsed $E_8$ theory, while the GUT singlet spectrum can not be embedded.

\begin{table}
	\centering
	\begin{tabular}{|c|c|}
		\hline
		Model	& spectrum embedded in\\\hline\hline
		\multicolumn{2}{|c|}{No $U(1)$ models}\\\hline\hline
		\cite{Garcia-Etxebarria:2014qua,Mayrhofer:2014haa}	 &	$\left\{ 2,2,2 \right\}_2$\\\hline
		\cite{Mayrhofer:2014haa}	 &	$\left\{ 2,2,2 \right\}_2$\\\hline
		\multicolumn{2}{|c|}{One $U(1)$ models}\\\hline\hline
		\cite{Braun:2013yti}	 &	$\left\{ 3,4,2 \right\}$\\\hline
		\cite{Borchmann:2013hta}, \cite{Kuntzler:2014ila} fiber type $I_5^{(01)}$&	  $\left\{ 3,3,2 \right\}$\\\hline
		\cite{Kuntzler:2014ila} fiber type $I_{5,ncnc}^{(01)}$&	  $\left\{ 3,3,2 \right\}$\\\hline
		\cite{Borchmann:2013hta}, \cite{Kuntzler:2014ila} fiber type $I_5^{(0|1)}$ &  $\left\{ 4,5,4 \right\}$ or $\mathbf{\left\{ 2,3,2 \right\}}$\\\hline
		\cite{Borchmann:2013hta}, \cite{Kuntzler:2014ila} fiber type $I_{5,nc}^{(0|1)}$ &  $\mathbf{\left\{ 2,3,2 \right\}}$\\\hline
		\cite{Borchmann:2013hta}, \cite{Kuntzler:2014ila} fiber type $I_{5,nc}^{(0||1)}$ & $\left\{ 3,4,3 \right\}$\\\hline
		\multicolumn{2}{|c|}{Two $U(1)$'s models}\\\hline\hline
		\cite{Mayrhofer:2012zy} $4-1$ split& $\mathbf{\left\{ 2,2,1 \right\}}$\\\hline
		\cite{Mayrhofer:2012zy} $3-2$ split& $\mathbf{\left\{ 2,3,2 \right\}}$\\\hline
		Top 1 & $\mathbf{\left\{ 3,5,6 \right\}}$\\\hline
		Top 2 & $\left\{ 5,8,12 \right\}$\\\hline
		Top 3 & $\left\{ 4,6,7 \right\}$\\\hline
		Top 4 & $\left\{ 4,6,8 \right\}$\\\hline
		\cite{Braun:2014qka} & $\left\{ 5,8,12 \right\}$\\\hline
		$I_5^{s(0|1||2)}$ $(2,2,2,0,0,0,0,0)$ & $\mathbf{\left\{ 3,4,4 \right\}}$, $\left\{ 4,6,7 \right\}$, $\left\{ 5,8,12 \right\} *$\\\hline
		$I_5^{s(0|1|2)}$ $(2,1,1,1,0,0,1,0)$ & $\mathbf{\left\{ 3,5,6  \right\}}$ \\\hline
		$I_5^{s(0|1||2)}$ $(2,1,1,1,0,0,1,0)$ & $\left\{ 5,8,12  \right\}$\\\hline
		$I_5^{s(1|0|2)}$ $(3,2,1,1,0,0,0,0)$ & $\left\{ 5,8,12  \right\}$\\\hline
		$I_5^{s(01|2)}$ $(3,2,1,1,0,0,0,0)$ & $\left\{ 4,6,8 \right\}$\\\hline
		$I_5^{s(0|12)}$ $(4,2,0,2,0,0,0,0)$ & Not embeddable \\\hline
		$I_5^{s(012)}$ $(5,2,0,2,0,0,0,0)$ & Not embeddable \\\hline
		$I_5^{s(01||2)}$ $(2,2,2,0,0,0,0,0)$ & $ \left\{ 4,6,7 \right\}$\\\hline
		$I_5^{s(0|1||2)}$ $(2,1,1,1,0,0,0,0)$ & $\mathbf{\left\{3,5,6 \right\}}$ *\\\hline
		$I_5^{s(01||2)}$ $(2,1,1,1,0,0,0,0)$ & $ \left\{ 4,6,7 \right\}$\\\hline
		$I_5^{s(1|0|2)}$ $(2,1,1,1,0,0,0,0)$ & $ \left\{ 5,8,12 \right\}$\\\hline
		$I_5^{s(0|2||1)}$ $(1,1,1,1,0,0,1,0)$ & $ \left\{ 5,8,12 \right\}$\\\hline
		$I_5^{s(0|1||2)}$ $(1,1,1,0,0,0,0,0)$ & No consistent way to turn off non-flat points.\\\hline
		\cite{Grimm:2013oga} 2 Fibrations & Any of the 2 $U(1)$ models\\\hline

	\end{tabular}
	\caption{Known models and the spectrum they are embeddable in. The two U(1) models come from \cite{Borchmann:2013hta} and \cite{Lawrie:2014uya}. An asterisk means that one needs to turn off the non-flat points to find an embedding. The models marked in bold have $SU(5)$ charged matter which is associated to the $E_8$ part of the tree, see figure \ref{modelpaths}, though all such embeddings, with the exception of $\mathbf{\left\{ 2,2,1 \right\}}$, require beyond $E_8$ singlets.}
	\label{tab:embed}
\end{table}

There are two models, constructed in \cite{Lawrie:2014uya}, which were not embeddable in our classification. They contain non-flat points but the analysis in appendix \ref{app:nf} shows that in principle, for a restricted class of bases of the fibration, it is possible to turn them off by an appropriate choice of fibration. This also turns off some of the matter curves but still the remaining spectrum is not embeddable. There are two features of these models which may be related to this property. The first is that they do not form complete networks but only partially complete ones (as defined in section \ref{sec:globfe8}), ie. there are $\f$ matter curves which do not have a $\un \f \fb$ coupling. If one attempts to restrict the fibration so as to turn off these $\f$ matter curves then also the single $\te$ matter curve must be turned off and there is no $E_6$ Yukawa point which places them outside our classification. They are the only models which have this feature. Therefore our classification still includes all of the 27 models which form complete flat networks.

The second interesting feature of the models is that they exhibit co-dimension three points, located on matter curves, where the discriminant potentially enhances in vanishing order but there is no coupling associated to the point. As an example, the discriminant for the model $I_5^{s(0|12)}$ takes the schematic form
\be
\Delta \sim \sigma_2 \Delta_5 w^5 + \Delta_6 w^6 + {\cal O}\left( w^7 \right) \;.
\ee
Here $\sigma_2=w=0$ corresponds to a $\f$-matter curve \cite{Lawrie:2014uya}. The component $\Delta_5$ is some function which does not vanish over $\sigma_2=0$. The unusual property is that 
\be
\left.\Delta_6 \right|_{\sigma_2=0} = s_{3,1} \tilde{\Delta}_6 \;,
\ee
where $s_{3,1}$ is some section of the fibration. However there is no intersection of matter curves at the locus $\sigma_2=s_{3,1}=w=0$. Now the vanishing order of the discriminant at this point can be either 6 or 7 depending on the vanishing order of $\sigma_2$ at this point. If $\sigma_2$ vanishes to order one, then the discriminant vanishes to order 6, as it does over the rest of the matter curve, and there is no enhancement. However if the vanishing order of $\sigma_2$ is higher then there is an enhancement of the vanishing order of the discriminant over this locus, but no known associated physics. It can be checked that it is not possible to turn off all such points where this feature occurs in the fibration consistently. We do not know if the fact that these models are not embeddable is related to this feature or not.\footnote{It is also interesting to note that this dependence of the vanishing order of the discriminant on the vanishing order of some sections occurs in other models in the literature. In particular for top 2 in \cite{Borchmann:2013hta} one finds this over the full matter curve $c_{2,1}=w=0$. This curve also happens to exhibit a non-flat point. It would be interesting to study this feature of fibrations further.}

\subsubsection{$SO(10)$ models}

The extension of the set of theories reached by Higgsing $E_8$ in the case of $SU(5)$ was based on the fact that the $SU(5)$ singlets coming from the adjoint of $E_8$ were not sufficient to form a complete network with the charged matter. As mentioned already, this is not the case for the higher GUT groups. Consider the next case of an $SO(10)$ GUT which arises from the breaking $E_8 \rightarrow SO(10) \times SU(4)$. Then the decomposition of the adjoint of $E_8$ is 
\be
{\bf 248} \rightarrow \left({\bf 45},{\bf 1} \right) \oplus \left({\bf 16},{\bf 4} \right) \oplus \left({\bf \overline{16}},{\bf \overline{4}} \right) \oplus \left({\bf 10},{\bf 6} \right) \oplus \left({\bf 1},{\bf 15} \right) \;.
\ee
In terms of charges of the Cartan of $SU(4)$ we have that the ${\bf 16}$ reps are associated to $t_i$, ${\bf 10}$ reps are associated to $t_i+t_j$, and the $SO(10)$ singlets are associated to $t_i-t_j$, with $i=1,..,4$ and $\sum_i t_i=0$. Then the analogue situation to the pairs of $\f$ and $\fb$ we can make for $SU(5)$ would be pairs of $\te$ representations here, but we see that for any such pair there is an appropriate singlet to make a cubic coupling.  

With this in mind it is interesting to consider $SO(10)$ F-theory geometries with Abelian sectors and their possible embedding in a Higgsed $E_8$.  We could expect that if the fact that the $SU(5)$ geometries were not embeddable in $E_8$ is attributed to the missing singlets this should not occur for the case of $SO(10)$. Let us consider then some example $SO(10)$ models. We construct these as the tops over the $P_{[1,1,2]}$ and $P_{[1,1,1]}$ fibrations (which are the most general ones for one and two $U(1)$s \cite{Morrison:2012ei,Borchmann:2013jwa,Cvetic:2013nia}) following \cite{Bouchard:2003bu} . Note that some aspects of the one $U(1)$ cases were studied already in \cite{Kuntzler:2014ila}. The elliptic fibre for the two cases are given by
\bea
P_{[1,1,2]} &=& w^2 + b_0 u^2 w + b_1 u v w + b_2 v^2 w - u \left(c_0 u^3 + c_1 u^2 v + c_2 u v^2 + c_3 v^3 \right) \;, \\
P_{[1,1,1]} &=& v w \left(c_1  w + c_2 v \right)  + u  \left(b_0 v^2 + b_1 v w  + b_2 w^2 \right) +  u^2 \left(d_0 v  + d_1 w  + d_2 u\right) \;.
\eea
In table \ref{tab:so10tops} we present the matter curves, their couplings and the non-flat points for the constructions. We did not calculate the $U(1)$ charges as we will see that they are not needed for our purposes. 
\begin{table}
	\centering
	\begin{tabular}{|c|c|c|c|c|}
		\hline
		Model & $\te$-matter	&  ${\bf 16}$-matter & ${\bf 16\; 16\; 10}$ coupling & Non-flat loci \\\hline\hline
		\multicolumn{5}{|c|}{$SO(10)\times U(1)$ models}\\\hline\hline
		Top 1 & $b_{2} \;, c_{1,2}$ & $c_{2,1}$ & $c_{2,1} \cap b_2$ & $c_{2,1} \cap c_{1,2}$ \\\hline
		Top 2 & $b_{0,2} \;, c_{3}$ & $c_{2,1}$ & $c_{2,1} \cap b_{0,2}$ & $c_{2,1} \cap c_{3}$ \\\hline
		Top 3 & $b_{2} c_{1,2} - b_{0,1} c_{3,1}$ & $b_{0,1}\;,b_2$ & $b_{0,1} \cap b_{2}\;, b_{0,1} \cap c_{1,2}$ & $c_{3,1} \cap b_2$ \\\hline
		Top 4 & $c_{3,1}$ & $b_{2}$ & - & $c_{3,1} \cap b_{2}$ \; , $b_{0,1}$ \\\hline
		Top 5 & $c_{1,2}$ & $b_{0,1}$ & $c_{1,2} \cap b_{0,1}$  & $b_{2}$ \\\hline\hline
		\multicolumn{5}{|c|}{$SO(10)\times U(1)\times U(1)$ models}\\\hline\hline
		Top 1 & $b_{0} \;, d_{1,1}$ & $d_{0}$ & $d_{1,1} \cap d_0$ & $d_{0} \cap b_{0}\;, c_{1,1}$ \\\hline
		Top 2 & $c_{2} \;, b_{2,2} d_0 - c_{1,1} d_{2,1}$ & $d_{0}\;,c_{1,1}$ & $d_{0} \cap c_2 \;, d_{0} \cap c_{1,1} $ & $d_{0} \cap d_{2,1}\;, c_{1,1} \cap c_2 \;, c_{1,1} \cap b_{2,2}$ \\\hline
		Top 3 & $b_{2,2} \;, c_{2}$ & $c_{1,1}$ & - & $c_{1,1} \cap b_{2,2}$\;,$c_{1,1} \cap c_2\;, d_{0}$ \\\hline
		Top 4 & $c_{2} \;, d_{2,1}$ & $d_{0}$ & $c_{2} \cap d_0$ & $d_{0} \cap d_{2,1}\;, c_{1,1}$ \\\hline
		Top 5 & $c_{2} \;, d_{0,1}$ & $b_{0}$ & - & $b_{0} \cap c_{2}\;,b_{0} \cap d_{0,1}\;, b_{2,1}$ \\\hline		
	\end{tabular}
	\caption{$SO(10)$ models based on the top constructions in \cite{Bouchard:2003bu}. We show the matter curves, the coupling points, and the non-flat loci. The notation $c_{i,j}$ is such that the second index denotes the vanishing order of the section in the coordinate defining the $SO(10)$ divisor.}
	\label{tab:so10tops}
\end{table}
There are two immediately apparent differences between the $SO(10)$ and $SU(5)$ constructions: Firstly the number of matter curves for $SO(10)$ models is very small, and secondly the non-flat loci are more prolific in the $SO(10)$ case. These two facts together imply that flat $SO(10)$ fibrations are very restricted. We see that apart from Top 3 for $SO(10) \times U(1)$ all the models are such that flatness implies a single $\te$ curve and a single ${\bf 16}$ curve with a coupling intersection.\footnote{The model based on Top 2 for $SO(10) \times U(1) \times U(1)$ requires a bit of analysis to show that there is no possibility to make it flat while keeping two ${\bf 16}$ curves, but it can be shown to not be possible by requiring effectiveness of the sections in the fibration in an analysis similar to those presented in appendix \ref{app:nf}.} Such a setup is easily embedded in a Higgsed $E_8$ theory. Top 3 for $SO(10) \times U(1)$  can lead to a flat model by setting $c_{3,1}$ to be a constant, which leads to a model with two ${\bf 16}$ matter curves, denoted ${\bf 16}^1$ and ${\bf 16}^2$, and a single $\te$ matter curve. The couplings present are ${\bf 16}^1\; {\bf 16}^1\; \te $ and  ${\bf 16}^1\; {\bf 16}^2\; \te $. This implies a one parameter family of possible charges for the curves. A one parameter family of charges satisfying the conditions for such gauge invariant couplings to be present can be reached by Higgsing $E_8$ by setting $t_1=t_2=t_3$. Therefore this model is also embeddable in a Higgsed $E_8$. The full set of models constructed therefore are all embeddable in a Higgsed $E_8$ theory, consistent with the fact that the singlet coming from the adjoint of $E_8$ are sufficient to form a complete network.

\subsection{Some Phenomenological Applications}
\label{sec:pheno}

The models constructed in section \ref{HiggsBeyondE8} open up phenomenological applications which were not considered in local model building based on the Higgsing of $E_8$, see \cite{Maharana:2012tu} for a review of this literature. We will not present a systematic study of the possible phenomenologically realistic models that can arise from this set of theories, but discuss instead a number of interesting aspects. One such aspect of the new models is that they admit discrete symmetries. It is therefore natural to identify the MSSM matter-parity with a $\mathbb{Z}_2$ factor. In table \ref{rparity} we list the possible embedding of matter parity in the models which also allows for a gauge invariant top quark Yukawa coupling. As an example consider the first model in table \ref{rparity} based on $\{3,4,3\}_2$. We consider taking all three generations to be supported on one matter curve. In this case the assignment of charges under the $U(1)\times \mathbb{Z}_2$ symmetry group is 
\be
Q\left(\te_{\mathrm{up}}\right) = 1_1 \;,\;\; Q\left(\fb_{\mathrm{down}}\right) = 2_1 \;,\;\; Q\left(\f_{\mathrm{H_u}}\right) = -2_0 \;\,\;\; Q\left(\fb_{\mathrm{H_d}}\right) = -3_0 \;.
\ee
With these charges the $\mu$-term can be induced through the vev of the singlet $Q\left(\un_{\mu}\right) = 5_0$, or through the F-term of its conjugate via the Giudice-Masiero mechanism \cite{Giudice:1988yz}. It is also possible to implement neutrino masses in this model in two ways. The first is through the see-saw mechanism by choosing the right-handed neutrino to be the singlet with charge $Q\left(\un_{\nu_R}\right) = 0_1$. Note that it naturally picks up a Majorana mass since it has a gauge invariant mass term. The other option is, following the mechanism in \cite{ArkaniHamed:2000bq}, to take $Q\left(\un_{\nu_R}\right) = -5_1$ which forbids a Dirac or Majorana mass but allows for the K\"ahler potential term $\left(\fb_{\mathrm{H_d}}\right)^{\dagger} \fb_{\mathrm{down}} \un_{\nu_R}$ which leads to the correct Neutrino mass scale after supersymmetry and electroweak symmetry breaking. 

\begin{table}
	\begin{center}
		\begin{tiny}
		\begin{tabular}{|c|c|c|c|c|c|c|c|}
			\hline
			& up Yukawa & down Yukawa & $\mu$-term & Dir. mass & D-5 PD & Maj. Mass & RHN\\
			& $\FU^0 \TM^1 \TM^1$ & $\FD^0 \FM^1 \TM^1$ & $\OO^0 \FM^1 \FU^1$ & $\FU^0 \FM^1 \ON^1$ & & $\OO^0 \ON^1 \ON^1$ & $(\FD^0)^\dagger \FM^1 \ON^1$\\
			\hline\hline
			$\{3,4,3\}_2$ & $\TM: (1)_1$ & $\FD: (-3)_0$ & $\OO: (5)_0$ & $\ON: (0)_1$ & $\times$ & $\checkmark$ & $\ON: (-5)_1$\\
			& $\FU: (-2)_0$ & $\FM: (2)_1$ & & & & &\\\hline
			$\{3,4,3\}_2$ & $\TM: (1)_1$ & $\FD: (2)_0$ & $\times$ & $\ON: (5)_1$ & $\checkmark$ & $\times$ & $\ON: (5)_1$\\
			& $\FU: (-2)_0$ & $\FM: (-3)_1$ & & & & &\\\hline\hline
			$\{4,5,5\}_2$ & $\TM: (2)_1$ & $\FD: (4)_0$ & $\times$ & $\ON: (10)_1$ & $\checkmark$ & $\times$ & $\ON: (10)_1$\\
			& $\FU: (-4)_0$ & $\FM: (-6)_1$ & & & & &\\\hline\hline
			$\{5,7,7\}_2 $ & $\TM: (1)_1$ & $\FD: (-3)_0$ & $\OO: (5)_0$ & $\ON: (0)_1$ & $\times$ & $\checkmark$ & $\ON: (-5)_1$\\
			& $\FU: (-2)_0$ & $\FM: (2)_1$ & & & & &\\\hline
			$\{5,7,7\}_2 $ & $\TM: (1)_1$ & $\FD: (2)_0$ & $\times$ & $\ON: (5)_1$ & $\checkmark$ & $\OO: (10)_0$ & $\ON: (5)_1$\\
			& $\FU: (-2)_0$ & $\FM: (-3)_1$ & & & & &\\\hline
			$\{5,7,7\}_2 $ &  $\TM: (1)_1$ & $\FD: (7)_0$ & $\OO: (-5)_0$ & $\ON: (10)_1$ & $\times$ & $\times$ & $\ON: (15)_1$\\
			& $\FU: (-2)_0$ & $\FM: (-8)_1$ & & & & &\\\hline
		\end{tabular}
		\end{tiny}
	\end{center}
	\label{rparity}
	\caption{Table showing the models which support a top quark Yukawa coupling and a $\mathbb{Z}_2$ symmetry with charges matching those of matter parity in the MSSM.}
\end{table}

A proposition made in \cite{Dudas:2009hu} is to associate different generations to different matter curves. In particular this potentially allows for an understanding of flavour physics of the Standard Model in terms of the additional $U(1)$ symmetries via the Froggatt-Nielsen mechanism \cite{Froggatt:1978nt}.\footnote{We note that the non-$E_8$ singlets play a nice role in the flavour physics model presented in \cite{Dudas:2009hu}. There was a problem in that model that the Guidice-Masiero mechanism for the $\mu$-term required coupling to a singlet which had charges beyond those of the $E_8$ ones. The new singlets however have the appropriate charges to play the role of the Guidice-Masiero field.}
We explored if the new theories beyond $E_8$ offer new opportunities for realising this idea. However we find no nice implementations of models where each generation is on a different curve and also dimension four proton decay operators are forbidden. The underlying reason is that essentially one needs to have a generation-universal $U(1)$ symmetry to control dimension four proton decay operators. Now since the single three-$U(1)$ model beyond $E_8$ does not enjoy a gauge-invariant rank one up-quark Yukawa matrix at tree level, realistic flavor physics should come from two-$U(1)$ models. Combined with one $U(1)$ being used for proton decay operators this does not leave sufficient freedom to create the full flavour structure. We do find models where two generations lie on the same curve which can partially reproduce realistic flavour physics but clearly not fully. Of course if one was to dismiss dimension four proton decay operators, say by considering high-scale supersymmetry breaking, then realistic flavour models can be found. We will not discuss this possibility further at this point though.

\section{Comments on Heterotic Duality}
\label{sec:hetd}

Some F-theory vacua are dual to compactifications of the Heterotic string. The general statement is that the Heterotic string compactified on a Calabi-Yau $(n+1)$-fold $X_{n+1}$ which is an elliptic fibration over a base $B_n$ is dual to F-theory on a Calabi-Yau $(n+2)$-fold $Y_{n+2}$ which is elliptically fibered over a base $\hat{B}_{n+1}$ that is itself a $\mathbb{P}^1$ fibration over the original Heterotic base $B_n$. Since the perturbative Heterotic string is based on a ten-dimensional $E_8\times E_8$ theory the role of $E_8$ as a group which is Higgsed down is apparent. Given the duality a natural question then arises regarding what are the Heterotic duals of the singlets lying outside $E_8$? In this section we study this question. Heterotic/F-theory duality is not so well understood for the compactifications that we are considering, in particular the presence of additional sections and these being four-dimensional rather than six-dimensional compactifications means that the duality is rather complicated (see however \cite{Choi:2012pr} for some work on this). This means that we will not be able to specifically identify the dual states on the Heterotic side but instead present a collection of results which may provide clues as to their nature. 

The geometry on the F-theory side maps to both bundle data and geometry on the Heterotic side. Specifically, if $w=0$ is the locus of the non-Abelian divisor, in this case carrying $SU(5)$, then we should consider a Weierstrass formulation of the fibration
\be
y^2 - x^3 = f x + g \;,
\ee
and expand the $f$ and $g$ coefficients in powers of $w$
\be
f = \sum_i f_i w^i \;,\;\; g = \sum_i g_i w^i \;. \label{fgexpan}
\ee
The prescription introduced in \cite{Morrison:1996pp} is that $f_i$ with $i=0,...,3$ and $g_i$ with $i=0,...,5$ encode the information on the bundle in the $E_8$ factor containing the $SU(5)$. The terms $f_4$ and $g_6$ encode the geometry of the CY. And the higher order terms encode information on the second $E_8$. We will study the geometry on the Heterotic side for the specific example models constructed in \cite{Borchmann:2013hta}. First though we can make some general statements regarding the bundle data.

\subsection{Bundle data}

We begin be reviewing how the bundle data is recovered in the familiar case of compactifications to six dimensions \cite{Morrison:1996pp,Bershadsky:1996nh}. In this case the base of the F-theory elliptic fibration must be a Hirzebruch surface $\mathbb{F}_n$ with base coordinate $z$. Then the coefficients of the expansions (\ref{fgexpan}) have to be functions of $z$ of degrees given by
\be
\mathrm{deg}\left( f_i \right)= 8 + n\left( 4 - i\right) \;, \;\;\; \mathrm{deg} \left( g_i \right) = 12 + n\left( 6 - i\right) \;.
\ee 
Such configurations are dual to Heterotic bundles with $12+n$ instantons in the first $E_8$ and $12-n$ instantons in the second $E_8$. The particular role of $E_8$ that we are interested in is as the group in which the instantons are embedded, which determines their moduli space. In particular if we want to preserve an $SU(5)_{GUT}$ so that the instantons must sit completely in $SU(5)_{\perp}$, then their moduli space is of dimension $36+5n$. More generally, and appropriately for generalising this configuration to four dimensions, the bundle associated to the instantons is described by $36+5n$ parameters which are the coefficients of 5 functions $b_i$ where $i=0,2,3,4,5$ with degrees $12+n-2i$.\footnote{The degrees of the functions lead to $13+9+7+5+3+5n=37+5n$ parameters but one must subtract an overall rescaling of all the functions which is a rescaling of $w$.} The 5 functions $b_i$ are then the data of the spectral cover description of the Heterotic bundle \cite{Friedman:1997yq}. Now the $f_i$ and $g_i$ are ten functions which, if generic, would encode $114+31n$ degrees of freedom, far more than $36+5n$. They are however not generic because they must be restricted so that we have a remaining $SU(5)$ singularity. This implies that they have relations which reduce their degrees of freedom to the correct number and we will come back to the precise form of these relations soon. 

Generalising this to four-dimensional compactifications the actual parameter counting will change, since now the $f_i$ and $g_i$ are functions on a surface rather than $\mathbb{P}_1$, see \cite{Friedman:1997yq} for the generalisation, but the overall logic remains the same. The bundle on the Heterotic side is still an $SU(5)$ spectral cover bundle which is specified by 5 functions $b_i$ of the appropriate degree in the base coordinates. These count deformations of the bundle inside of $E_8$. We now come to the map between the $f_i$ and $g_i$ and the $b_i$. If we restrict to fibrations which can be written in Tate form then this relation can be deduced by mapping to Weierstrass form thereby getting a relation between the $a_{i,j}$ and the $f_i$ and $g_i$. Then one can attempt to write the ten functions $f_i$ and $g_i$ in terms of five functions $b_i$ which will themselves be functions of the $a_{i,j}$. An analysis of this was performed in \cite{Bershadsky:1996nh} focusing on six-dimensional compactifications with no additional sections. We find the general solution to this problem is
\bea
f_0 &=& -1/48 b_5^4 \;,\;\;
f_1 = -1/12 b_5^2 b_4 \;,\;\;
f_2 = -1/12 (b_4^2 - 6 b_5 b_3) \;,\;\;
f_3 = 1/24 b_2 \;, \\
g_0 &=& 1/864 b_5^6 \;,\;\;
g_1 = 1/144 b_5^4 b_4 \;,\;\;
g_2 = 1/72 b_5^2 (b_4^2 - 3 b_5 b_3) \;,\nonumber \\
g_3 &=& -1/864 (3 b_2 b_5^2 - 8 b_4^3 + 72 b_5 b_4 b_3) \;, \;
g_4 =  1/144 (-b_2 b_4 + 36 b_3^2) + \Delta g_{4}\;,\nonumber \\
g_5 &=& 1/288 b_0 \;, \nonumber
\eea
where the functions which appear are given explicitly by
\bea
b_5 &=& a_{1,0} \;,\nonumber \\
b_4 &=& 2 a_{2,1} + a_{1,1} b_5   \;,\nonumber \\
b_3 &=& a_{3,2}  + 1/12 (-a_{1,1}^2 b_5 - 4 a_{2,2} b_5 - 2 a_{1,2} b_5^2) \;,\nonumber \\
b_2 &=& 24 a_{4,3} + (a_{1,1}^3 b_5 + 4 a_{1,1} a_{2,2} b_5 + 12 a_{3,3} b_5 - 
    4 a_{2,3} b_5^2 - 2 a_{1,3} b_5^3 - 2 a_{1,1}^2 b_4 \nonumber \\
		& &- 8 a_{2,2} b_4 - 4 a_{1,2} b_5 b_4 + 12 a_{1,1} b_3) \;,\nonumber \\
b_0 &=& 288 a_{6,5} - \left(a_{1,1}^2 b_2 + 4 a_{2,2} b_2 - 12 a_{1,1}^2 a_{3,3} b_5 - 
    48 a_{2,2} a_{3,3} b_5 + 2 a_{1,2} b_2 b_5   \right. \nonumber \\
		& &- 12 a_{1,2} a_{3,3} b_5^2 + 12 a_{1,1} a_{3,4} b_5^2 + 24 a_{4,5} b_5^2 + a_{1,1}^2 a_{1,3} b_5^3 + 
    4 a_{1,3} a_{2,2} b_5^3 + 12 a_{3,5} b_5^3 \nonumber \\
		& &- 2 a_{1,1} a_{1,4} b_5^4 - 4 a_{2,5} b_5^4 - 2 a_{1,5} b_5^5 + 24 a_{1,1} a_{3,3} b_4 + 48 a_{4,4} b_4  \nonumber \\
		& &+ 2 a_{1,1}^2 a_{1,2} b_5 b_4 + 
    8 a_{1,2} a_{2,2} b_5 b_4 + 24 a_{3,4} b_5 b_4 - 8 a_{1,1} a_{1,3} b_5^2 b_4  \nonumber \\
		& &  - 16 a_{2,4} b_5^2 b_4 - 8 a_{1,4} b_5^3 b_4 - 8 a_{1,1} a_{1,2} b_4^2 - 
    16 a_{2,3} b_4^2 - 8 a_{1,3} b_5 b_4^2  \nonumber \\
		& & \left. - 144 a_{3,3} b_3 + 24 a_{1,1} a_{1,2} b_5 b_3 + 48 a_{2,3} b_5 b_3 + 36 a_{1,3} b_5^2 b_3 + 24 a_{1,2} b_4 b_3\right) \;, \nonumber \\
\Delta g_4 &=& 1/576 b_5^2 \left(a_{1,1}^4 + 8 a_{1,1}^2 a_{2,2} + 16 a_{2,2}^2 - 24 a_{1,1} a_{3,3} - 48 a_{4,4} + 
   2 a_{1,1}^2 a_{1,2} b_5 + 8 a_{1,2} a_{2,2} b_5 \right. \nonumber \\
	 & & - 24 a_{3,4} b_5 + 2 a_{1,2}^2 b_5^2 + 
   4 a_{1,1} a_{1,3} b_5^2 + 8 a_{2,4} b_5^2 + 4 a_{1,4} b_5^3 + 8 a_{1,1} a_{1,2} b_4 + \nonumber \\
	 & & \left.16 a_{2,3} b_4 + 8 a_{1,3} b_5 b_4 - 24 a_{1,2} b_3\right) \;.
\eea
The important point is that it is not possible to write the $f_i$ and $g_i$ in terms of just the $b_i$ but there is a left-over piece $\Delta g_4$. Note that if we restrict the Tate coefficients to only their leading components in the expansion in $w$, so setting $a_{i,n}=0$ for $n>i-1$, then we find $\Delta g_4=0$ and $b_i \sim a_{6-i,5-i}$. So the leading order coefficients in the Tate form can be mapped to the degrees of freedom associated to a bundle embedded in $E_8$ through the spectral cover. However the sub-leading coefficients can not all vanish else the dual heterotic geometry would be singular (the discriminant of the dual elliptic fibration would vanish identically). We therefore see that applying the Heterotic/F-theory duality prescription reveals more degrees of freedom than would be associated to $E_8$, those encoded in $\Delta g_4$, which are associated to the sub-leading powers in $w$ in the Tate model. Note that this is consistent with the results of section \ref{ExaBeyondE8} where we showed that Higgsing away from $E_8$ involves a deformation of also the sub-leading terms, in that case $a_{4,4}$. 

It is interesting to note that the inability to write the $f_i$ and $g_i$ in terms of the degrees of freedom associated to the spectral cover occurs for the first time for a spectral cover of $SU(5)$. In the sense that for $SU(2)$ we have that $f$ must vanish to order 3 and $g$ to order 5 so the data is encoded in two functions, $f_3$ and $g_5$ which match the spectral cover data. For $SU(3)$ we have $f_3$, $g_4$ and $g_5$, which match onto the 3 functions in the spectral cover. For $SU(4)$ we have $f_2$, $f_3$, $g_3$, $g_4$ and $g_5$. Now there are 5 functions on the F-theory side but only 4 functions are needed to specify an $SU(4)$ bundle. However if we take $f$ and $g$ to originate from a Tate form, then it is simple to find a single relation between the functions $f_2^3 \sim g_3^2$, and therefore there are only 4 independent functions, matching the spectral cover degrees of freedom. Another way to see this is that $\Delta g_4$ vanished over the locus $b_5=0$ which enhances $SU(5)$ to $SO(10)$. The first discrepancy therefore appears for $SU(5)$ matching the observation already raised that the singlets extending the adjoint of $E_8$ are only necessary for a complete network in the case of an $SU(5)$ GUT, while for $SO(10)$ and higher this is not so. It is tempting therefore to associate the additional degrees of freedom in $\Delta g_4$ with the additional singlet fields present on the F-theory side. However a specific map would require a more detailed understanding of the duality. 

\subsection{Geometry data}

Since the additional singlets on the F-theory side are not embeddable in the adjoint of $E_8$ it is natural to associate their heterotic duals with non-perturbative states.\footnote{It appears unlikely that they are perturbative states associated with the second $E_8$ factor since at finite string coupling the two $E_8$ branes are separated and states charged under both become massive.} The appearance of non-perturbative gauge symmetries in the Heterotic string have been studied through F-theory duality extensively in six dimensions, see \cite{Friedman:1997yq,Aspinwall:1997ye} for early papers, and there have been some studies of four-dimensional cases, for recent work which includes a literature overview see \cite{Anderson:2014gla}. In the six-dimensional case non-perturbative Heterotic gauge symmetries appeared when small instantons/bundle degenerations, or colliding M5-M9 branes, were placed on a singularity in the geometry. On the F-theory side they appeared as non-minimal singularities which required a blow-up of the base geometry. There are straightforward four-dimensional generalisations of this phenomenon to curves of bundle degenerations \cite{Rajesh:1998ik,Diaconescu:1999it}. In four-dimensions also co-dimension three non-minimal loci/bundle degenerations can occur \cite{Candelas:1997eh}. The Heterotic origin of the additional singlets however remains unclear to us, in this section we simply study some potential clues. In particular we are interested in a possible correlation between non-$E_8$ singlets and singularities in the dual heterotic geometry. 

Our first analysis was of the tops constructions of \cite{Borchmann:2013hta}. We relegate the details of the analysis to appendix \ref{app:hettops}, and here present an outline of the calculation and state the results. We consider these fibrations on a general base which is appropriate for Heterotic duality, ie. such that it is a ${\mathbb P}^1$ fibration. The top fibrations exhibit non-flat points and so the first step is to restrict them such that these non-flat loci are absent. We find that this is possible, but restrictive, for tops 2, 3 and 4, while for top 1 it is not possible. Next we analysed the dual Heterotic geometry by studying the discriminant from the $f_4$ and $g_6$ coefficients.  We find that there are singular loci in the Heterotic geometry over points in the base where in the F-theory geometry some singlet matter intersects the GUT brane. The particular singlets responsible for the singularities, and the detail of the singularities are shown in table \ref{sing_loc}. Note that they are all $SU(2)$ singularities, similar to those found on the F-theory side over curves supporting the singlets. 

\begin{table}
	\begin{center}
		\begin{tabular}{c|c|c|c|c|c|c}
		& singular singlets & $f_4$ & $g_6$ & $\Delta_{\text{het}}$  & Singularity type\\\hline\hline
		top 2 & $\mathbf{1}_3$ & $1$ & $2$ & $3$ &  $SU(2)$\\\hline
		top 3 & $\mathbf{1}_1$ & $2$ & $2$ & $4$ &  $SU(2)$\\\hline
		top 4 & $\mathbf{1}_3$ & $0$ & $0$ & $2$ &  $SU(2)$\\
		& $\mathbf{1}_5$ & $2$ & $2$ & $4$ &  $SU(2)$\\
		\end{tabular}
		\caption{Table showing the vanishing order of $f_4$, $g_6$ and $\Delta_{\text{het}}$ over points associated to the intersection of singlets with the GUT divisor in F-Theory.}
		\label{sing_loc}
	\end{center}
\end{table}

For tops 2 and 3 it is possible to restrict the fibration further such that the singular loci in table \ref{sing_loc} are absent and the heterotic dual geometry is smooth. In that case also the matter spectrum is reduced, since some of the curves get turned off, and the resulting spectrum can be embedded in a Higgsed $E_8$ theory. If we do not restrict to a smooth fibration we find that the spectrum for both top 2 and 3 can not be embedded in $E_8$. These results are consistent with an association of the heterotic duals of the singlets beyond $E_8$ with singularities. 

However, such an identification does not apparently work for a different type of fibration constructed in \cite{Mayrhofer:2012zy}, the 3-2 factorised Tate fibration presented in section \ref{ExaBeyondE8}. There the non-$E_8$ singlet was associated to the point $\delta=\beta=w=0$, and it can be checked that the Heterotic dual geometry does not exhibit a singularity over this point.\footnote{This is true for generic coefficients for the subleading, in $w$, terms of the Tate sections. It is possible to find non-generic choices where there is a singularity present at that point. However we have no arguments for why that particular choice must be imposed.} We therefore do not find a complete correlation between geometric properties of the heterotic dual and the F-theory singlets extending $E_8$. For the top models the singlets beyond $E_8$ can be associated to singular loci in the heterotic geometry, but this does not appear to hold for the factorised Tate model. 

\section{Summary}
\label{sec:summary}

In this work we studied the interplay between global F-theory GUTs and the group $E_8$. In particular we defined an extension of the set of theories that can be reached by a standard Higgsing of $E_8$. The extension amounts to introducing additional GUT singlets which do not arise from the adjoint representation of $E_8$ and Higgsing using these singlets to reach new theories. We gave an explicit geometric construction of this process where a global F-theory model, the so called 3-2 Factorised Tate model, included such an additional singlet and could be deformed to a different fibration which amount to Higgsing by this additional singlet. We then classified the full set of possible theories that could be reached by this process, extending the 6 Higgsed $E_8$ model types by an additional 20. We presented the full set of representations and Abelian charges for these theories. 

We went on to compare this extended set of theories with explicit fibrations constructed in the literature. In total we considered 44 fibrations: 30 resolved $SU(5)$ fibrations listed in table \ref{tab:embed}, 10 $SO(10)$ fibrations in table \ref{tab:so10tops}, and 4 more given as factorised Tate models in \cite{Mayrhofer:2012zy} for which no resolution was presented. Of these, one could not be made flat and two more did not form complete networks (as defined in section \ref{sec:globfe8}) and the remaining 37 resolved fibrations could all be embedded into our extended set of theories. Of these only 11 could also be embedded into a Higgsed $E_8$ theory: the 10 $SO(10)$ models and the $4-1$ factorised Tate. We noted that this could correspond to the fact that, in contrast to $SU(5)$, in Higgsing $E_8$ to $SO(10)$ no new singlets need to be introduced in order for all the possible cubic couplings between fields to be present.

We also made some comments regarding the heterotic duals of the F-theory fibrations which lie outside a Higgsed $E_8$ theory. We noted that sometimes, but not always, there can be a correlation between singlets outside of $E_8$ and singularities on the heterotic dual geometry. Also that the F-theory data which should encode bundle data on the heterotic side contains more degrees of freedom than an $E_8$ spectral cover bundle construction. However we could not identify explicitly the heterotic dual origin for the singlet fields extending $E_8$ that were introduced in this work on the F-theory side.

The work presented is very much just an initial exploration of the relation between $E_8$ and global F-theory GUTs. There are many possible directions to explore along this path. Most straightforwardly is to continue to check for more F-theory geometries which are embeddable in $E_8$, which in our extended set of theories, and which in neither. The more ambitious goal, which this could be a step towards, is a geometric derivation of the implications of the existence of a co-dimension three $E_6$ point for the full theory on the GUT surface. It is likely that the intersection structure of the matter curves would play a crucial role in such a geometric understanding. In this work we restricted to complete networks which means that the fibration is generic enough that all the cubic couplings which could be present by gauge invariance are present. A next logical step would be an understand of what happens in less generic cases where the fibration is such that some intersection points are missing. Along the same lines, an incorporation of further more complicated effects like fluxes and gluing modes into the question of the relation with $E_8$ would be interesting.

A general direction of future work which relates to some of the themes explored in this paper is a better understanding of Heterotic/F-theory duality in four dimensions and with additional sections. The fact that this duality has not been studied to any great detail was one of the reasons we found identifying the Heterotic states dual to the F-theory singlets difficult. It would be very nice to have a better understanding of this duality and the relation to both perturbative line bundle models \cite{Anderson:2011ns,Anderson:2012yf,Anderson:2013xka} and extensions of the perturbative $E_8$ symmetry.

The new models arising from Higgsing beyond $E_8$ in F-theory have potential applications to phenomenological model building and therefore it would be worthwhile to construct their associated geometries explicitly. Something which we noted is the presence of additional discrete symmetries. Another point, raised in \cite{Braun:2013yti}, is that additional matter representations can relax anomaly cancellation constraints on hypercharge flux as studied in \cite{Marsano:2010sq,Palti:2012dd,Mayrhofer:2013ara} leading to more possible realisations of this mechanism.

\noindent
{\bf Acknowledgments}
\newline
\noindent
We would like to thank Timo Weigand for initial collaboration on this project and for many valuable discussions. We also thank Lara Anderson, James Gray, Stefan Groot Nibbelink, Hirotaka Hayashi, Craig Lawrie, Ling Lin, Christoph Mayrhofer, Sakura Schafer-Nameki and Oskar Till for valuable discussions and explanations. The work of FB and EP is supported by the Heidelberg Graduate School for Fundamental Physics. 

\appendix

\section{Embedding in the presence of non-flat points}
\label{app:nf}

Some of the two $U(1)$ models studied in \cite{Lawrie:2014uya} are not directly embeddable in the tree. They however contain non-flat points that once turned off also turn off matter curves. These models are labeled by their Kodaira fiber $I$ and the two sets
\begin{gather}
		(n_1,n_2,n_3,n_4,n_5,n_6,n_7,n_8)\nonumber\\
	\left[ d_{2,n_1},d_{0,n_2},b_{0,n_3},d_{1,n_4},b_{1,n_5},c_{2,n_6},b_{2,n_7},c_{1,n_8}\right].
\end{gather}
The integers $n_i$ denote the leading non-vanishing order of the Tate model coefficients, while the terms in square brackets define a specialisation of the Tate form coefficients. The homology classes of the coefficients are combinations of three classes on the base $B_3$ denoted $\bar{\mc{K}}$, $\alpha$, and $\beta$ and are given in table \ref{eF/classes}. 
\begin{table}
	\begin{center}
		\begin{tabular}{c|c|c|c|c|c|c|c}
 			$b_0$ & $b_1$ & $b_2$ & $c_1$ & $c_2$ & $d_0$ & $d_1$ & $d_2$\\\hline
 			$\alpha - \beta + \bar{\mc{K}}$ & $\bar{\mc{K}}$ & $- \alpha + \beta + \bar{\mc{K}}$ & $-\alpha + \bar{\mc{K}}$ & $- \beta + \bar{\mc{K}}$ & $\alpha + \bar{\mc{K}}$ & $\beta + \bar{\mc{K}}$ & $\alpha + \beta + \bar{\mc{K}}$\\
		\end{tabular}
		\caption{The classes of the sections in the fibration of \cite{Lawrie:2014uya,Borchmann:2013jwa,Cvetic:2013nia}. $\bar{\mc{K}}$ is the anti-canonical class of the base $B_3$.}
		\label{eF/classes}
	\end{center}
\end{table}
There are five a priori non-embeddable models:
\begin{enumerate}
	\item The first model is  
		\begin{equation}
			I^{s(0|1||2)}_5~:\qquad
			\bigg\{	
			\begin{array}{c}  
				(2,2,2,0,0,0,0,0)\\
				\left[-,-,-,\sigma_2\sigma_5,\sigma_2\sigma_4+\sigma_3\sigma_5,\sigma_3\sigma_4,\sigma_1\sigma_2,\sigma_1\sigma_3\right]
			\end{array}
			\bigg\}.
		\end{equation}
		It has non flat points at the loci $\left\{ \sigma_2=\sigma_3=0 \right\}$ and $\left\{\sigma_4=\sigma_5=0\right\}$. From table \ref{eF/classes}, we can read off the classes of the sections $\sigma_i$ to be:
		\begin{align*}
			[d_{2,2}]&=\alpha+\beta+\bar{\mathcal{K}}-2\omega\\
			[d_{0,2}]&=\alpha+\bar{\mathcal{K}}-2\omega\\
			[b_{0,2}]&=\alpha-\beta+\bar{\mathcal{K}}-2\omega\\
			[d_1]&=[\sigma_2]+[\sigma_5]=\beta+\bar{\mathcal{K}}\\
			[b_1]&=[\sigma_2]+[\sigma_4]=[\sigma_3]+[\sigma_5]=\bar{\mathcal{K}}\\
			[c_2]&=[\sigma_3]+[\sigma_4]=-\beta+\bar{\mathcal{K}}\\
			[b_2]&=[\sigma_1]+[\sigma_2]=-\alpha+\beta+\bar{\mathcal{K}}\\
			[c_1]&=[\sigma_1]+[\sigma_3]=-\alpha+\bar{\mathcal{K}}
		\end{align*}
		There are thus four possibilities to turn them off:
		\begin{enumerate}
			\item $\left[ \sigma_2 \right]=\left[ \sigma_4 \right]=0$: This implies that the anti-canonical bundle $\bar{\mathcal{K}}=0$, which is inconsistent.
			\item $\left[ \sigma_3 \right]=\left[ \sigma_5 \right]=0$: Same case as the previous one, hence inconsistent.
			\item $\left[ \sigma_2 \right]=\left[ \sigma_5 \right]=0$: This implies $[d_1]=-\alpha$ and $[b_2]=\alpha$. At least one of those classes are not effective, which is inconsistent.
			\item $\left[ \sigma_3 \right]=\left[ \sigma_4 \right]=0$. In that case, we must turn off two $\mathbf{\bar{5}}$ curves. The resulting spectrum is then embeddable in several models (see table \ref{tab:embed}).
		\end{enumerate}

	\item Model 
		\begin{equation}
			I^{s(0|1||2)}_5~:\qquad
			\bigg\{	
			\begin{array}{c}  
				(2,1,1,1,0,0,0,0)\\
				\left[-,\sigma_1\xi_3,\sigma_1\xi_2,-,\sigma_4\xi_3,\sigma_4\xi_2,\xi_3\xi_4,\xi_2\xi_4\right]
			\end{array}
			\bigg\}
		\end{equation}
		has three non-flat points at $\left\{ \sigma_1=\sigma_4=0 \right\}$, $\left\{ \sigma_4=\xi_4=0 \right\}$ and $\left\{\xi_2=\xi_3=0\right\}$. Using a similar reasoning as before, one finds that the only consistent possibility to turn off these points is to set at least $\left[ \xi_4 \right]$ trivial, turning off a $\mathbf{\bar{5}}$ that then allow an embedding in a $\left\{ 3,5,6 \right\}$ model.
	
	\item 
		\begin{equation}
			I^{s(1|02)}_5~:\qquad
			\bigg\{	
			\begin{array}{c}  
				(4,2,0,2,0,0,0,0)\\
				\left[-,-,\sigma_3\sigma_4,-,\sigma_2\sigma_4+\sigma_3\sigma_5,\sigma_1\sigma_3,\sigma_2\sigma_5,\sigma_1\sigma_2\right]
			\end{array}
			\bigg\}
		\end{equation}
		has non flat points at the loci $\left\{ \sigma_2=\sigma_3=0 \right\}$ and $\left\{\sigma_4=\sigma_5=0\right\}$. There are therefore four consistent ways to turn off the non flat points. The first is to set $[\sigma_2]=[\sigma_5]=0$. This constraints the classes to $\beta=\alpha-\bar{\mathcal{K}}\leq0$, $\omega\leq\alpha/2$. The second possibility is to set $[\sigma_3]=0=[\sigma_4]$. This leads to the same constraints as before, with the role of $\alpha$ and $\beta$ reversed. The two remaining possibilities lead to a vanishing anti-canonical class, which is inconsistent. 
		
		We however find that even with a reduced spectrum, there is still no possible embedding into the tree.
	\item 
		\begin{equation}
			I^{s(012)}_5~:\qquad
			\bigg\{	
			\begin{array}{c}  
			(5,2,0,2,0,0,0,0)\\
			{[}-,\sigma_1\sigma_2,\sigma_2\sigma_5,\sigma_1\sigma_3,\sigma_2\sigma_4+\sigma_3\sigma_5,-,\sigma_3\sigma_4,-]
			\end{array}
			\bigg\}
		\end{equation}
		is similar to the previous one: It has non-flat points at the same loci, and there are two consistent ways to turn off the non-flat points. Either one sets the classes $[\sigma_2]=0=[\sigma_5]$. The classes are then constrained to $\alpha\leq0$, $\beta=\alpha+\bar{\mathcal{K}}\geq 5\omega/2$. The other consistent possibility is to set $[\sigma_3]=0=[\sigma_4]$. This gives rise to the same constraints on the classes, with the role of $\alpha$ and $\beta$ reversed.

		As for the previous case, we find no possible embedding in the tree.
	\item The last case,
		\begin{equation}
			I^{s(0|1||2)}_5~:\qquad
			\bigg\{	
			\begin{array}{c}  
				(1,1,1,0,0,0,0,0)\\
				{[}\xi_3\delta_3\delta_4,\delta_4(\delta_3\xi_2+\delta_2\xi_3),\xi_2\delta_2\delta_4,\xi_3\delta_1\delta_4,\\
				\qquad\qquad\delta_1(\delta_2\xi_3+\delta_3\xi_2),\delta_1\delta_2\xi_2,\sigma_1\xi_3,\sigma_1\xi_2{]}
			\end{array}
			\bigg\}
		\end{equation}
		has five non flat points:
		\begin{gather*}
			\left\{ \delta_1=\delta_2=0 \right\}\qquad\left\{ \delta_1=\delta_4=0 \right\}\\
			\left\{ \delta_1=\sigma_1=0 \right\}\qquad\left\{ \delta_2=\delta_3=0 \right\}\\
			\left\{ \xi_2=\xi_3=0 \right\}
		\end{gather*}
		We find that there is no consistent way to turn them off by setting classes of the different sections to zero.
\end{enumerate}

\section{Heterotic duality and $SU(5)\times U(1)\times U(1)$ tops}
\label{app:hettops}

We begin by reviewing the necessary elements of the constructions in \cite{Borchmann:2013hta}. The elliptic fibrations that we will use are written as a hypersurface in $\mathbb P^2$ 
\be
 P_T = \tv\, \tw (c_1\,  \tw + c_2\, \tv )  + \tu\, (b_0\, \tv^2 + b_1\, \tv\, \tw  + b_2\, \tw^2) +  \tu^2 (d_0\, \tv  + d_1\, \tw  + d_2\, \tu).
\label{eq:hyper1}
\ee
This hypersurface is singular and can be resolved through two blow-ups whose exceptional divisors are associated to the two sections responsible for the $U(1)\times U(1)$ gauge symmetry. The homology classes of sections appearing in (\ref{eq:hyper1}) are combinations of three classes on the base $B_3$ denoted $\bar{\mc{K}}$, $\alpha$, and $\beta$ and are given in table \ref{eF/classes}. In order to induce a further $SU(5)$ singularity we require certain vanishing orders for the $b_i$, $c_i$ and $d_i$ in the GUT divisor $w$. The possibilities are classified as the tops on this fibration which give 4 different models \cite{Borchmann:2013hta}. 

We are interested in studying the heterotic duals of these models which means we can assume that the base $B_3$ is a $\mathbb P^1$ fibration over a base $B_2$ which means that the anti-canonical class can be written as
\be
\bar{\mc{K}} = \mathrm{c}_1\left(B_2 \right) + 2 w + t \;,
\ee
where $w$ is the is the divisor supporting the $SU(5)_{GUT}$ group and $t$ is some class specifying the $\mathbb P^1$ fibration. 

The four top fibrations studied in \cite{Borchmann:2013hta} had co-dimension three points where the fibration was non-flat and the fibre jumped in dimension. As mentioned previously such loci are associated to tensionless strings from wrapped M5-branes which we should avoid in thinking about an embedding in $E_8$. Since these are co-dimension 3 points it may be that on a specific base they would just be absent, but if we wish to keep the analysis general and only study the fibration structure the way to guarantee their absence is to restrict the $b_i$, $c_i$ and $d_i$ sections in such a way as to forbid the non-flat points. 
Before we consider a particular top model, let us outline the general procedure we are going to follow. First we expand the line bundles $\alpha$ and $\beta$ into a piece depending on the base $B_2$ and one depending on $w$
\be
	\alpha = \alpha_{B_2} + a_w w \;,\;\;
	\beta = \beta_{B_2} + b_w w \label{eff_nfp/expansion} \;,\;\;
	\hspace{13pt} a_w, b_w \in \mathbb{Z} \;.
\ee
For a specific base $B_3$ we could further expand the first piece into the generators of the Mori cone of $B_2$. However we wish to consider generic $\mathbb P^1$ fibrations here and so can not do this. We can nevertheless derive some general restrictions: Since $w$ is independent from the classes coming from the $B_2$ piece we can consider it separately. Effectiveness of the classes of $b_i$, $c_i$ and $d_i$ will then give a set of inequalities acting on the $w$ coefficients $a_w$ and $b_w$. Of course such an analysis only gives a necessary constraint rather than a sufficient one, as we do not consider the effectiveness in the expansion of generators of $B_2$. The non-flat points are determined as an intersection of two curve classes and we turn the non-flat points off by setting one of the curve classes trivial in homology. This will then give an additional equality to be satisfied. 

As an example consider top 2 from \cite{Borchmann:2013hta}. In this model the leading vanishing orders of the sections in the coordinate $w$ normal to the divisor are given by:
\begin{align}
	b_0 = b_{0,3} w^3 \;,\;\;
	c_2 = b_{2,1} w \;,\;\;
	d_0 = d_{0,2} w^2 \;,\;\;
	d_2 = d_{2,1} w \;. \label{eff_nfp/vanish2}
\end{align}
Top 2 has a non-flat point at $\{b_1 = c_{2,1} = 0\}$, which we turn off by setting
\be
	[c_{2,1}] = \bar{\mc{K}} -w -\beta = 0 \;,
\ee
since we want to keep the $\te$ matter curve at $b_1=0$. This therefore fixes $b_w = 1$. 
Now requiring that all the $b_i$, $c_i$ and $d_i$ are effective also uniquely fixes $a_w = 2$.
The results of applying the same analysis to the other top models can be found in table \ref{eff_nfp/top2eff_nfp}.
\begin{table}
	\begin{center}
		\begin{tabular}{c|c|c|c|c}
			& top 1 & top 2 & top 3 & top 4\\\hline
 			non-flat point & $\{b_1 = d_1 = 0\}$ & $\{b_1 = c_{2,1} = 0\}$ & $\{b_1 = c_1 = 0\}$ & $\{b_1 = b_{0,1} = 0\}$\\\hline
			restriction on & cannot be & $a_w = 2$ & $a_w = 2$ & $a_w = -1$\\
			$a_w$, $b_w$ & satisfied & $b_w = 1$ & $b_w = 0$ & $b_w = 0$\\
		\end{tabular}
		\caption{Constraints on $a_w$ and $b_w$ from the effectiveness of the $k_{i,j}$ and turning off the non-flat points.}
		\label{eff_nfp/top2eff_nfp}
	\end{center}
\end{table}
In summary, for the models top 2--4 there is a unique choice of the $w$-coefficients of the line bundles $\alpha$, $\beta$ such that all sections are effective and the non-flat point is turned off. In contrast, there is no possibility to do this for top 1, that is it does not allow for a heterotic dual in terms of the geometries considered.

Having restricted the fibrations to be flat on the F-theory side we calculated the discriminant of the fibration of the Heterotic dual geometry (constructed from the $f_4$ and $g_6$ coefficients of the F-theory fibrations). We find that there are singularities in the Heterotic geometry over the points where in the F-theory dual some of the singlets intersect the GUT brane. In table \ref{sing_loc} we present the singularities and which singlet intersection points they are associated to. 

For tops 2 and 3 it is possible to restrict the fibration so that the singularities are absent. The restriction is shown in table \ref{sing_choices}. Restricting the fibration in this way also turns off some of the GUT singlets and charged matter on the F-theory side. For both top 2 and top 3 we find that before turning off the singular heterotic loci it is not possible to embed the matter (including the singlets) spectrum in $E_8$, while after restricting the fibration so that the Heterotic dual is smooth the resulting spectrum can be embedded in $E_8$. This embedding is given in table \ref{appemb}. 

\begin{table}
	\begin{center}
		\begin{tabular}{c|cc|c}
			& & constr. on coupling & turned off curves\\\hline\hline
			top 2 & ($\times$) & $\alpha = 2 c_1(B_2) + 2t + 3w$\\
			& ($\checkmark$) & $\alpha = c_1(B_2) +t +2w$ & $\mathbf{1}_{(1)}$,$\mathbf{1}_{(3)}$,$\mathbf{1}_{(5)}$,$\mathbf{5}_{(1)}$,$\mathbf{5}_{(2)}$\\\hline
			top 3 & ($\times$) & $\beta = 2 c_1(B_2) + 2t + 2 w$\\
			& ($\checkmark$) & $\beta = c_1(B_2) +t$ & $\mathbf{1}_{(1)}$,$\mathbf{1}_{(3)}$,$\mathbf{1}_{(5)}$,$\mathbf{5}_{(2)}$\\\hline
			top 4 & ($\times$) & $\alpha = c_1(B_2) +t + w$\\
			& ($\times$) & $2c_1(B_2) +2t +3w = 0$\\
			& & $\alpha = -w$\\\hline
		\end{tabular}
		\caption{Different choices to turn off the singular singlets. The $\checkmark$ and $\times$ symbol indicate whether a particular choice agrees with the constraints derived from turning off the non-flat point and having effective sections. Note that there is a unique choice to do so in top 2 and 3 and no such choice in top 4. The last column which curves are turned off by these restrictions.}
		\label{sing_choices}
	\end{center}
\end{table}

\begin{table}
	\begin{center}
		\begin{tabular}{|c|c|c|}
			\hline
			& top 2 & top 3\\\hline\hline
			$U(1)_1$ & $-t^1 +4t^2 -t^3 -t^4 -t^5$ & $t^1 -4t^2 -4t^3 +6t^4 +t^5$\\\hline
			$U(1)_2$ & $-2t^1 +3t^2 +3t^3 -2t^4 -2t^5$ & $t^1 +t^2 -4t^3 +t^4 +t^5$\\\hline\hline
			$\mathbf{10}$ & $t_1$ & $t_1$\\\hline\hline
			$\bar{\mathbf{5}}_A$ & $t_2 +t_3$ & $t_2 +t_5$\\\hline
			$\bar{\mathbf{5}}_B$ & $t_1 +t_2$ & $t_3 +t_4$\\\hline
			$\bar{\mathbf{5}}_C$ & $t_1 +t_4$ & $t_3 +t_5$\\\hline
			$\bar{\mathbf{5}}_D$ & $t_3 +t_4$ & $t_2 +t_4$\\\hline
			$\mathbf{1}_A$ & $t_2 -t_3$ & $t_1 -t_2$\\\hline
			$\mathbf{1}_B$ & $t_1 -t_2$ & $t_3 -t_1$\\\hline
			$\mathbf{1}_C$ & $t_3 -t_1$ & $t_2 -t_3$\\\hline
			\hline Higgsing & $t_1 = t_4 = t_5$ & $t_1 = t_2 = t_4$\\\hline
		\end{tabular}
	\end{center}
	\caption{Two example embeddings of matter and relevant singlet spectrum in $E_8$ for top 2 and 3 once the heterotic geometry is restricted to be smooth. Note that in these embeddings the singlets giving rise to singularities in the heterotic geometries are not embeddable.}
	\label{appemb}
\end{table}

\end{document}